\documentclass[pre,twocolumn,twoside,showpacs,byrevtex,superscriptaddress]{revtex4}

\usepackage{currfile}
\usepackage{url}
\usepackage{graphicx,epsfig,verbatim,enumerate}
\usepackage{amssymb,amsmath}
\usepackage{ifthen}
\newboolean{twocolswitch}

\usepackage{xcolor}
\usepackage{color}

\definecolor{lightgrey}{rgb}{0.5,0.5,0.5}

\usepackage{rotating}
\usepackage{longtable}
\usepackage{array}

\newcommand{\sindex}[1]{}
\newcommand{\nindex}[1]{}

\newcommand{\www}[1]{\url{#1}}

\newcommand{\Req}[1]{Eq.~(\ref{#1})}

\newcommand{\PreserveBackslash}[1]{\let\temp=\\#1\let\\=\temp}
\newcommand{\PBS}[1]{\let\temp=\\#1\let\\=\temp}

\newcommand{\pinstA}{\rm inst}

\newcommand{\popB}{\rm pop}

\newcommand{\MLEsinaionezerogamma}{2.02}
\newcommand{\MLEsinaionezerogammaconf}{0.10}

\newcommand{\MLEEinsteinonezerogamma}{1.80}
\newcommand{\MLEEinsteinonezerogammaconf}{0.06}

\newcommand{\MLEUVMonezerogamma}{1.81}
\newcommand{\MLEUVMonezerogammaconf}{0.05}

\newcommand{\MLEechozeroninegamma}{1.73}
\newcommand{\MLEechozeroninegammaconf}{0.15}

\newcommand{\MLEflynntheateronezerogamma}{2.09}
\newcommand{\MLEflynntheateronezerogammaconf}{0.05}

\newcommand{\MLEunitedwayonezerogamma}{2.47}
\newcommand{\MLEunitedwayonezerogammaconf}{0.09}

\setboolean{twocolswitch}{true}

\newcommand{\revtexonly}[1]{#1}
\newcommand{\plainlatexonly}[1]{}

\begin{document}

\title{
  Collective Philanthropy: Describing and Modeling the Ecology of Giving

}

\author{
\firstname{William L.}
\surname{Gottesman}
}
\email{billgottesman@gmail.com}

\affiliation{
  Computational Story Lab,
  University of Vermont,
  Burlington,
  VT, 05401
}

\author{
\firstname{Andrew James}
\surname{Reagan}
}
\email{andrew.reagan@uvm.edu}

\affiliation{
  Department of Mathematics and Statistics,
  Center for Complex Systems,
  \&
  the Vermont Advanced Computing Core,
  University of Vermont,
  Burlington,
  VT, 05401
}

\affiliation{
  Computational Story Lab,
  University of Vermont,
  Burlington,
  VT, 05401
}

\author{
\firstname{Peter Sheridan}
\surname{Dodds}
}
\email{peter.dodds@uvm.edu}

\affiliation{
  Department of Mathematics and Statistics,
  Center for Complex Systems,
  \&
  the Vermont Advanced Computing Core,
  University of Vermont,
  Burlington,
  VT, 05401
}

\affiliation{
  Computational Story Lab,
  University of Vermont,
  Burlington,
  VT, 05401
}

\date{\today}

\begin{abstract}
  Reflective of income and wealth distributions,
philanthropic gifting appears to follow an approximate power-law size distribution
as measured by the size of gifts received by individual institutions.
We explore the ecology of gifting by analysing 
data sets of individual gifts for a diverse group of institutions
dedicated to education, medicine, art, public support, and religion.
We find that the detailed forms of gift-size distributions differ across 
but are relatively constant within charity categories.
We construct a model for how a donor's income affects their giving preferences
in different charity categories, offering a mechanistic
explanation for variations in institutional gift-size distributions.
We discuss how knowledge of gift-sized distributions 
may be used to assess an institution's gift-giving profile,
to help set fundraising goals, 
and 
to design an institution-specific giving pyramid.
  
\end{abstract}

\maketitle

\section{Introduction}
\label{sec:phdist.introduction}

The scope and health of philanthropic institutions contribute
substantively to the cultural and economic well-being of a great
diversity of societal institutions.
Between 1970 and 2010 Americans gave approximately 2\% of their
disposable income to philanthropic causes \cite{givingusa2011a}.
The distribution of income in the United States and many other
countries has long been described by various heavy tailed
distributions including power law, log-normal, Boltzman, and
combinations thereof~\cite{clementi2005power,nirei2007two,wu2011a}.
Similar distributions have been found in the size of gifts to
charitable causes~\cite{chen2009a}.

\begin{figure*}[tbp!]
  \centering
  \includegraphics[width=\textwidth]{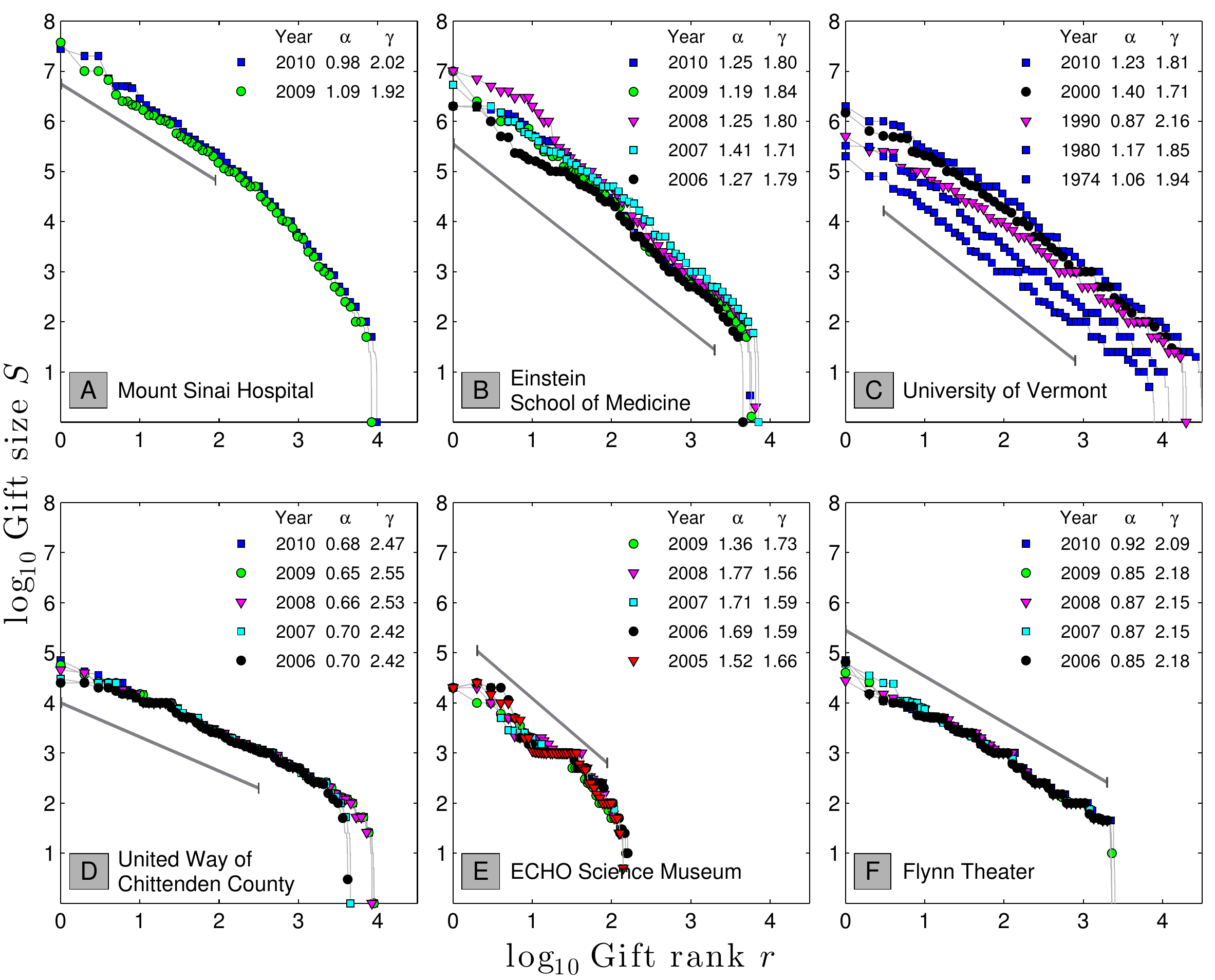}
 \caption{ 
    Gift size distributions for a range of institutions.
    The reported $\alpha$ and $\gamma$ were fitted to the region
indicated by solid grey line, and the 95\% CI of this fit, as well as
year for which the fit is plotted, are included for each organization.
    The ranges over which the data were fit was chosen empirically; other approaches were found to be inconsistent (see Supplementary).
    \textbf{A.} Health Care: Mt. Sinai Hospital, 2010 had $\gamma=\MLEsinaionezerogamma \pm \MLEsinaionezerogammaconf$.
    \textbf{B.} Higher Education (Medical): Einstein School of Medicine, 2010 had $\gamma=\MLEEinsteinonezerogamma \pm \MLEEinsteinonezerogammaconf$.
    \textbf{C.} Higher Education (General): University of Vermont, 2010 had $\gamma=\MLEUVMonezerogamma \pm \MLEUVMonezerogammaconf$.
    \textbf{D.} Combined Purpose: United Way, 2010 had $\gamma=\MLEunitedwayonezerogamma \pm \MLEunitedwayonezerogammaconf$.
    \textbf{E.} Cultural: ECHO Aquarium and Science Center, 2009 had $\gamma=\MLEechozeroninegamma \pm \MLEechozeroninegammaconf$.
    \textbf{F.} Performing Arts: Flynn Theater, Burlington VT, 2010 had $\gamma=\MLEflynntheateronezerogamma \pm \MLEflynntheateronezerogammaconf$.
    Later in Fig.~\ref{fig:phdist.OZstory}, we show similar data
    for an anonymous religious institution.
    Dates and amounts of all contributions were collected over time
    periods ranging from 2 years to 37 years.  
    United Way and Mt.~Sinai
    Hospital were the only organizations able to ensure that annual donor
    gift amounts reflected that year's total of donations by a single
    donor, rather than individually posting multiple gifts made by a
    single donor during that year.    
  }
  \label{fig:phdist.figphdists001}
\end{figure*}

Here, our aims are to 
(1) examine empirical data for an approximate
power-law size distribution model of philanthropic behavior; 
(2) describe a general
mathematical model for philanthropic gifting in a manner that gives
greater insight into how different organizations raise money and how
individuals choose the amounts of their gifts; 
and 
(3) explore the
usefulness of our findings on current fundraising
practices~\cite{pierpoint1998a}.
We have chosen the power law distributions
for the sake of simplicity and to aid development
of a primitive model describing heavy-tailed gifting behavior
capable of addressing basic questions about philanthropy.
We wish to emphasize that we do not claim that gift size distributions
are perfectly described as 'true' power laws generated by
some underlying mechanism(s) not yet elucidated.
Rather, we use power law
approximations---linear approximations in logarithmic coordinates---to gain some traction in our description
and to provide a way to carry out some idealized analysis,
fully appreciating the appromixate nature of our work.
Larger, much more comprehensive data sets will certainly
advance our understanding beyond what we have been able
to achieve here.

As a foundation for our investigations,
we have constructed a data set 
spanning a wide range of institutional categories.
We obtained anonymous gift data for a total of six institutions:
\begin{itemize}
\item 
  two educational institutions:
  University of Vermont, Burlington, VT, 
  and Albert Einstein Medical School,
  Bronx, NY, 
\item 
  one health care institution: Mt.~Sinai Hospital in
  Manhattan, NY,
\item 
  one combined purpose organization: United Way of
  Chittenden County, VT, 
\item 
  one local cultural and educational 
  organization: ECHO Science Center in Burlington, VT, 
\item 
  and one arts center: Flynn Theater in Burlington, VT.
\end{itemize}

Our central characterization of gift-size distributions 
will be through measuring power-law exponents for the 
paired statistics of Zipf distributions~\cite{zipf1949a} 
and gift-size frequency distributions, and we explain
both now.  
First, the Zipf distribution for a list of gifts is generated
by ranking gifts in order of descending monetary size.
Writing gift size as $S$ and gift rank as $r$, an ideal
Zipf distribution obeys:
\begin{equation}
  S \sim r^{-\alpha},
  \label{eq:phdist.zipfdistribution}
\end{equation}
where we will call $\alpha$ the Zipf exponent.
Alternately, we can compile a gift-size frequency distribution:
for each gift size $S$, we record the number of such gifts $N(S)$.
Again for an ideal system, we would observe
\begin{equation}
  N(S) \sim S^{-\gamma}.
  \label{eq:phdist.freqdistribution}
\end{equation}
Both views have their merits:
Zipf distributions follow a very natural construction
and are simple to interpret,
while power-law size distributions most clearly represent
a system's probabilistic behavior.
Later in our analyses, we will consider the probability
density $P(S)$, the normalized version of $N(S)$.

\begin{figure*}[tbp!]
  \includegraphics[width=\textwidth]{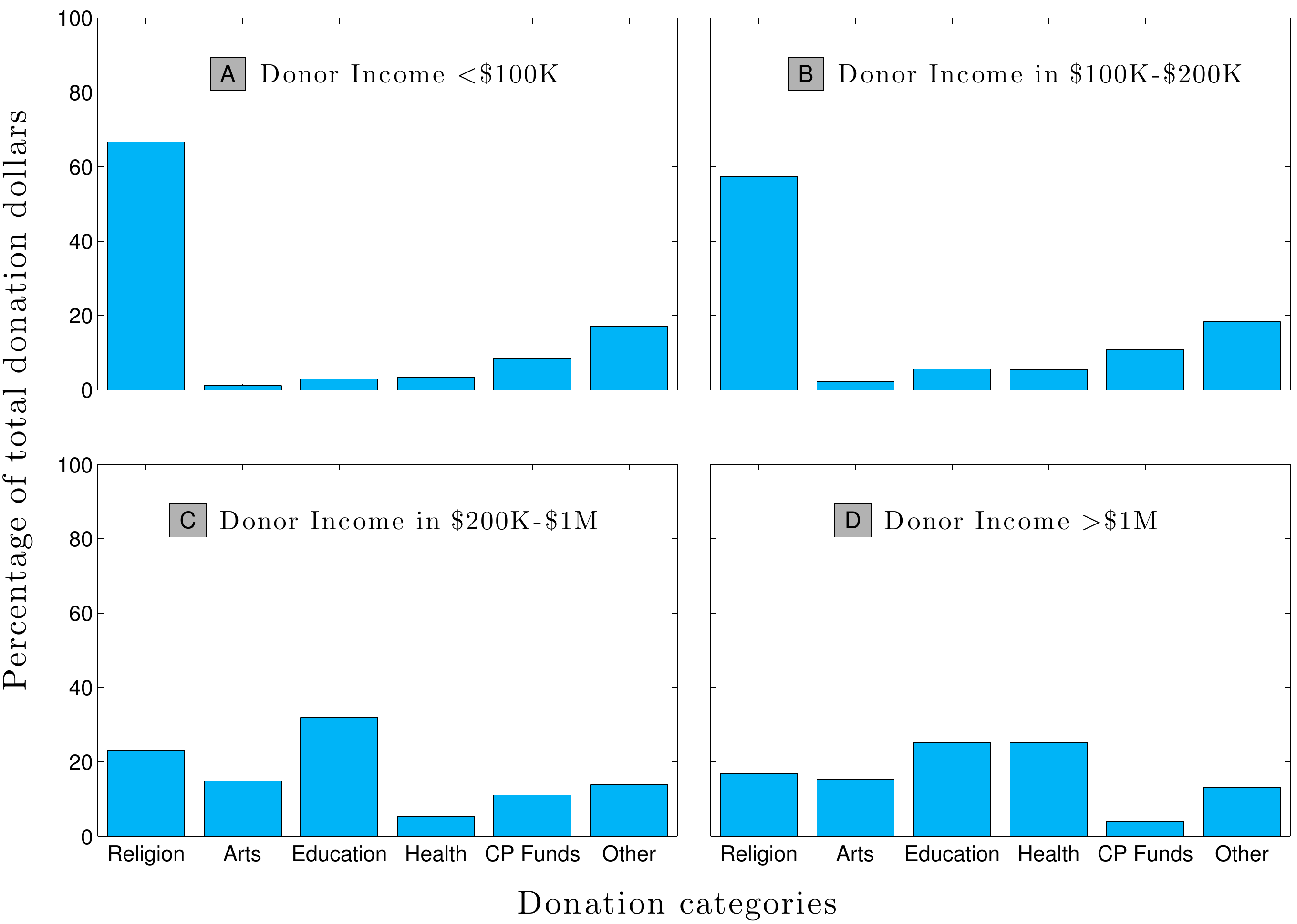}
  \caption{ 
    2005 data showing how donors of different income groups
    distribute their charitable giving.
    For example, on average donors earning less than \$100,000 chose
    to direct 67\% of their total giving to religious causes,
    panel \textbf{A}, but donors earning more than one million dollars chose
    instead to direct 17\%, panel \textbf{D}.
    CP Funds stands for Combined Purpose Funds.
  }
  \label{fig:phdist.Wealth-Giving-dist3a}
\end{figure*}

\begin{figure}[tbp!]
  \includegraphics[width=0.495\textwidth]{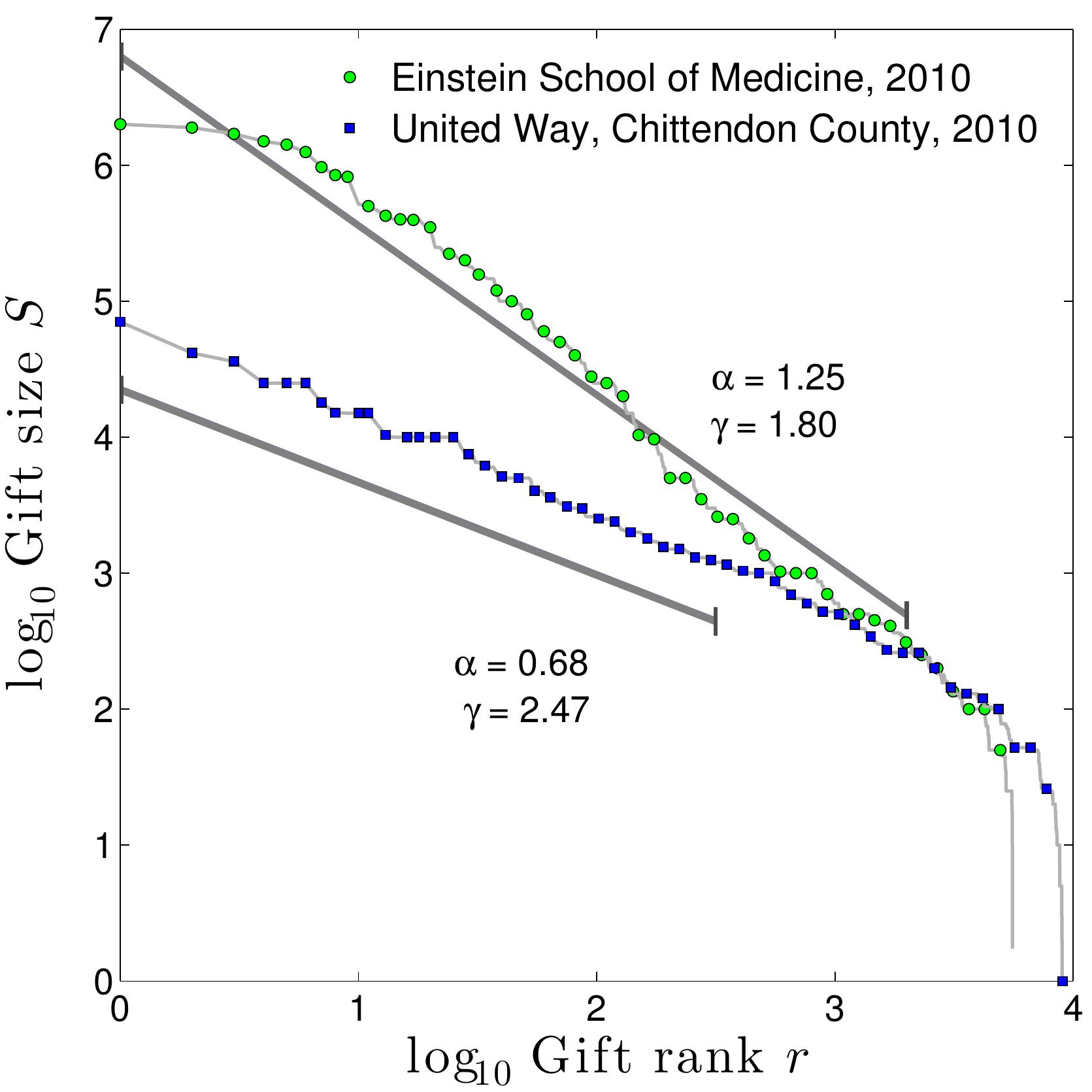}
  \caption{
    Comparison of 2010 giving to two organizations with a similar number of donors, and similarly sized smaller gifts.
    Despite these similarities, The Albert Einstein School of Medicine was able to attract top ranked gifts that were approximately 30 times larger than those of the United Way of Chittenden County, and raise 10 times the total, because Einstein enjoyed a substantially lower $\gamma$ than United Way.}
    \label{fig:phdist.gammacomparison}
\end{figure}

\begin{figure}[tbp!]
  \centering
  \includegraphics[width=0.495\textwidth]{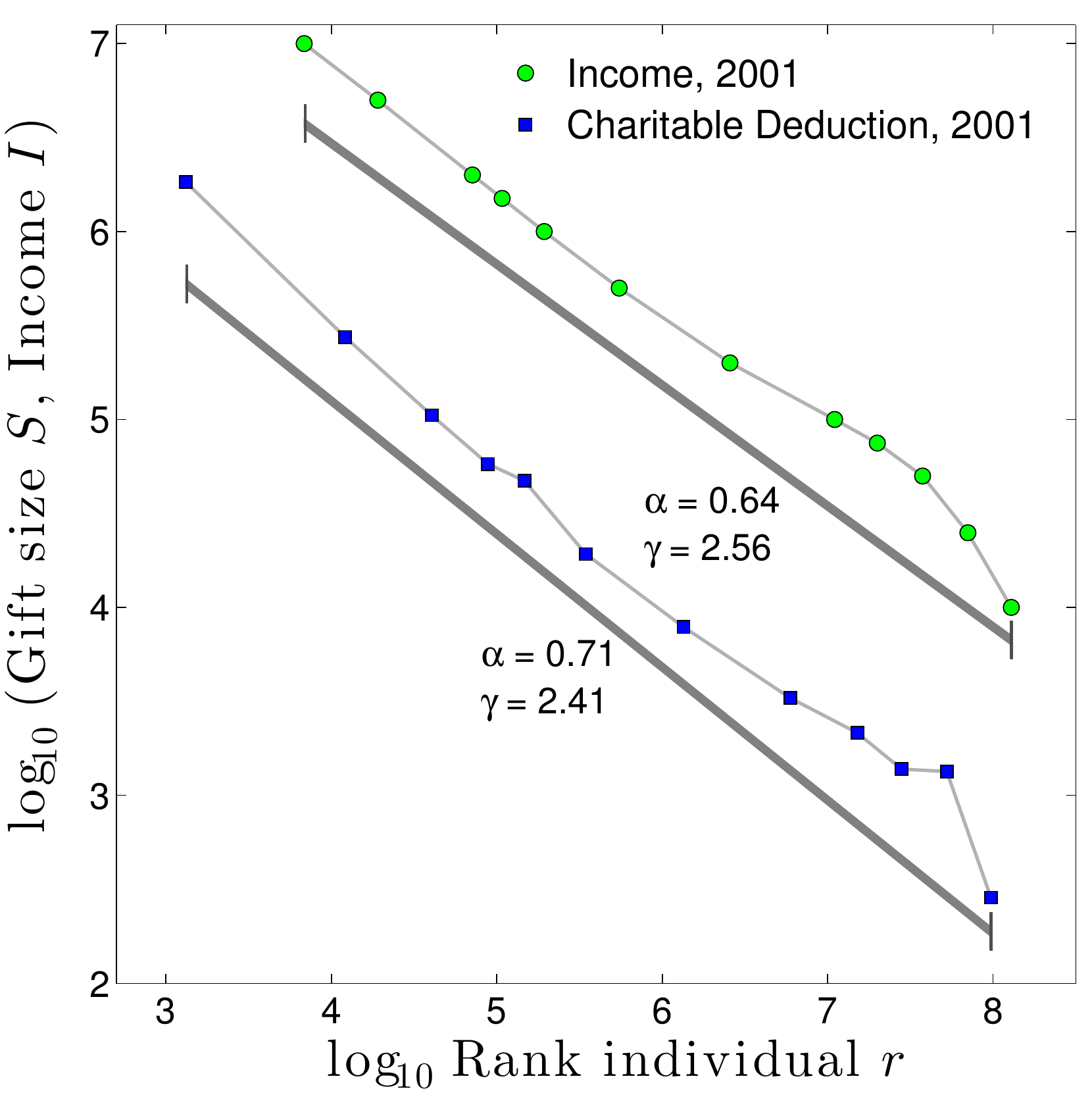}
  \caption{ 
    Data from IRS 2001 tax returns for personal income and
    charitable deductions.
    On average, people claimed charitable deductions at a rate of
    2.9\% of their income.
    The top 0.15\% of tax filers gave at a higher rate averaging
    4.8\%, resulting in a $\gamma$ for charitable deductions slightly
    lower (2.41) that that for income (2.56).
    Fits were computed using linear regression in log-log space, after
    attempts to use maximum likelihood methods failed due to finite
    size bias.
    Since income was reported as bin average, we found the rank of the
    individual with that average by assuming a power law distribution
    within each bin with $\gamma$ equal to the fit for the whole
    distribution.
    This procedure was bootstrapped (as the new individual ranks
    changes the whole distribution $\gamma$) until convergence of
    $\gamma$ within $10^{-3}$.
  }
  \label{fig:phdist.income}
\end{figure}

\begin{figure}[tbp!]
  \centering
  \includegraphics[width=0.495\textwidth]{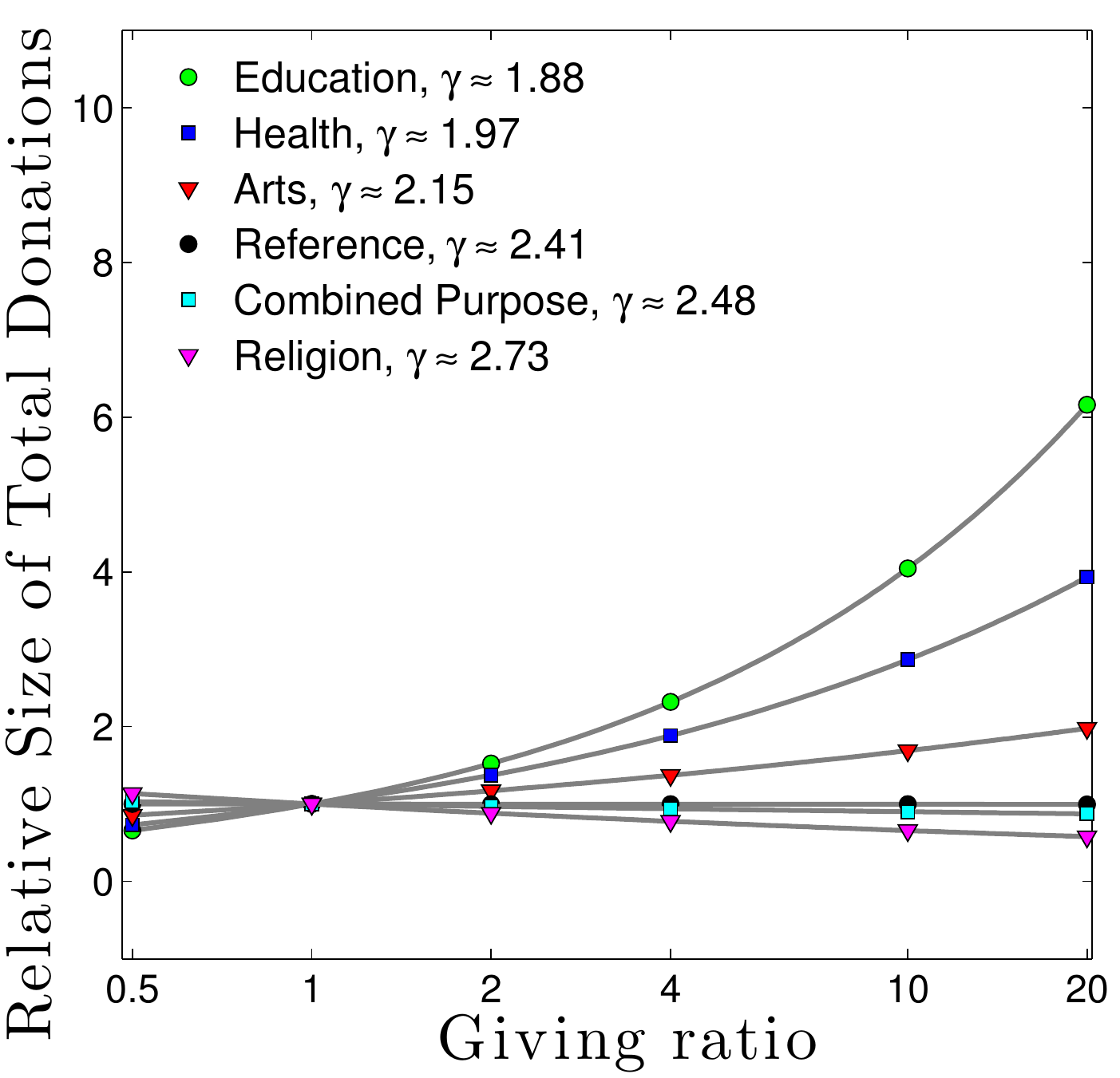}
  \caption{ 
    Examples of multipliers as a function of total donations
    and institutional categories as calculated by Equation 8.
    The X-axis measures the size of an individual's total donations
    relative to that of the index case of size 1.
    We calculated values of $\gamma$ from the average of all years of
    data shown in Figs. 1 and 13C.
    Education $\gamma$ is an average of $\gamma$'s from University of
    Vermont and Albert Einstein School of Medicine; Health $\gamma$ is
    from Mt. Sinai Hospital, Arts $\gamma$ is from Flynn Theater,
    Combined Purpose $\gamma$ is from United Way of Chittenden County,
    Reference $\gamma$ is from IRS 2001 charitable deductions Fig. 4,
    and Religion $\gamma$ is from Fig. 13C.
    Our model breaks down at extreme high and extreme low incomes
    where the multiplier could calculate a gift that would exceed
    100\% of that person’s total charitable giving.
  }
  \label{fig:phdist.transformingratio}
\end{figure}

\begin{figure*}[tbp!]
  \includegraphics[width=0.995\textwidth]{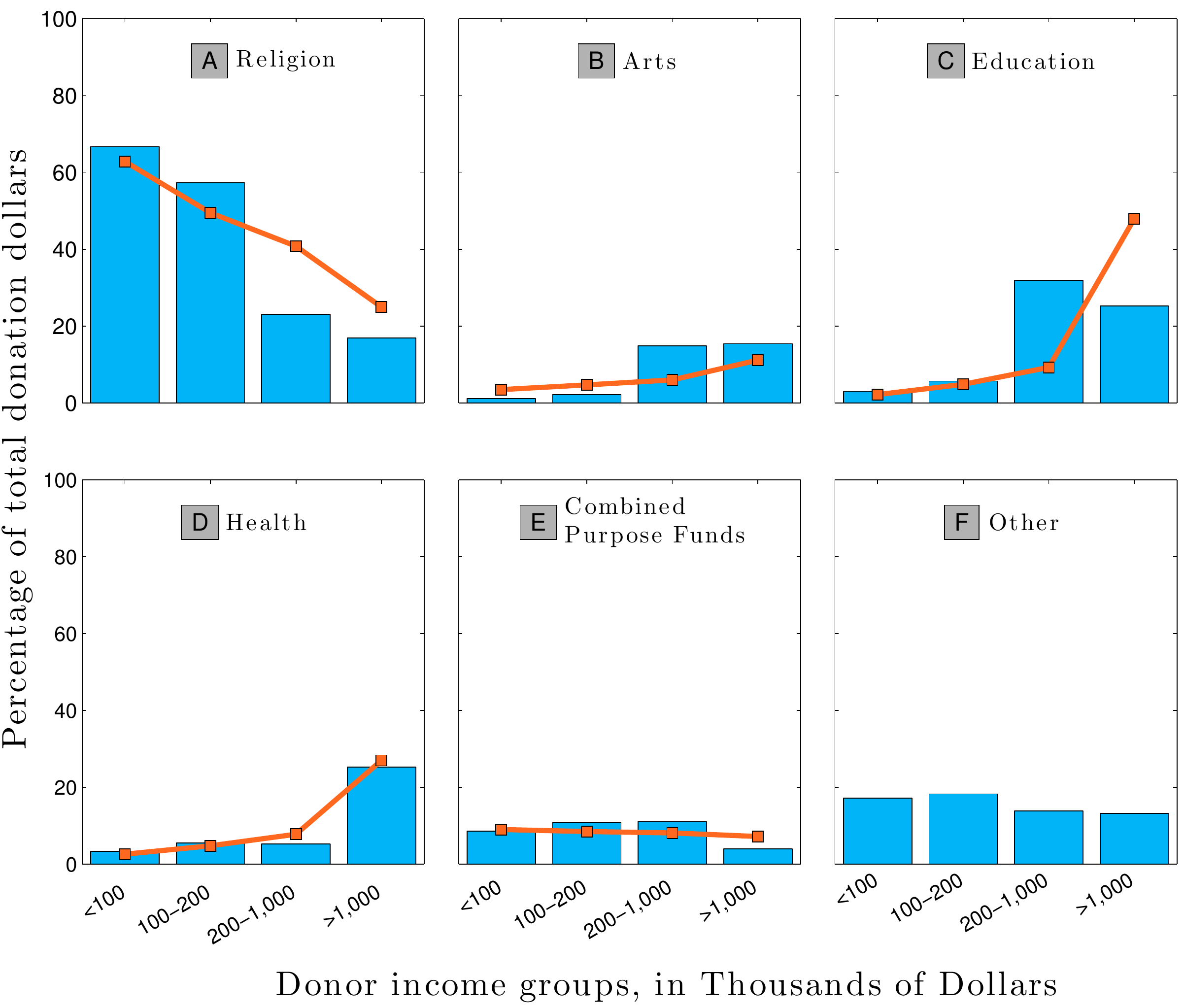}
  \caption{ 
    Allocation of donor giving choices as a function of
    income.
    Columns represent 2005 donor survey data [9].
    Connected squares represent our model of multipliers calculated
    from the values of $\gamma$ described in Figure
    \ref{fig:phdist.transformingratio}.
    The multiplier model agrees qualitatively with the donor survey
    data. 
  }
  \label{fig:phdist.Wealth-Giving-dist3b}
\end{figure*}

We can show that the two distributions are related 
by considering the complementary cumulative 
distribution function, the number of
gifts of at least size $S$: $N_{\ge}(S)$.
We see that $N_{\ge}(S)$ is equivalent to the rank of $S$,
meaning 
$N_{\ge}(S) = r \sim S^{-1/\alpha}$ using~\Req{eq:phdist.zipfdistribution}.
A simple calculation starting from the size frequency
distribution gives
$N_{\ge}(S) = \sum_{S'=S}^{S_{\rm max}} N(S') \sim S^{-(\gamma-1)}$.
The exponents are therefore connected as: 
\begin{equation}
  \alpha = \frac{1}{\gamma-1}.
  \label{eq:phdist.alphagamma}
\end{equation}
Empirically, a typical range for $\alpha$ is 1/2 to 1 with these limits 
corresponding to $3$ and $2$ for $\gamma$.
For $2 < \gamma < 3$, we have the `statistics of surprise':
gifts are typically small but the variance is very large being
dominated by the largest gifts.
If $\gamma < 2$, we have an even more extreme circumstance
of the average gift size being typically large as well.

In what follows, we will generally present figures showing Zipf's distribution.
We will estimate Zipf's $\alpha$ with a Maximum 
Likelihood (ML) approach~\cite{clauset2009b}, and then
determine $\gamma$ using~\Req{eq:phdist.alphagamma}.
We provide details of these calculations including
comparisons to other potential distributions in
the Supplementary Material
(see Fig.~\ref{fig:KSvsD} and 
Tabs.~\ref{table:fitstats} and
\ref{table:fittest}.)
In spite of our choice for figures and for
the purposes of analysis, we will prefer to 
describe our findings using the gift-size distribution 
exponent $\gamma$, though occasionally we will
use Zipf's $\alpha$ when more convenient.

Due to the real-world nature of our data sets,
our measurements are necessarily not exact.
Further, we are not stating that all 
philanthropic gift-size distributions possess idealized power-law tails.   
Power-law statistics are notoriously
difficult to estimate and, moreover, convincingly showing that a power law
even applies is itself a fraught endeavor~\cite{clauset2009b,dodds2001d}.
Nevertheless, assuming approximate power-law tails is reasonable
and gives us a serviceable diagnostic tool for building and challenging
our descriptions and theory.

We present our work in the following manner.
In Sec.~\ref{sec:phdist.empricalfindings}, 
we present and give an overall analysis of our six philanthropic-giving data sets.
In Sec.~\ref{sec:phdist.mechanism}, we propose an explanation
for the variation in gift-size distributions across institutions,
based on the gift-giving preferences of individual donors.
In Sec.~\ref{sec:phdist.recommendations}
we give recommendations for fundraisers
concerning 
the so-called `top 12 rule',
fundraising pyramids,
organization fundraising capacity,
and
data collection.
We provide concluding remarks
in Sec.~\ref{sec:phdist.phdist.conclusion}.

\section{Basic empirical findings} 
\label{sec:phdist.empricalfindings}

In Fig.~\ref{fig:phdist.figphdists001}, we show gift-size Zipf distributions
for our six organizations, organized by calendar year for a total of 27 distributions.
For each year's distribution for each institution, we estimate $\alpha$ and $\gamma$,
indicating the fitting region with a solid grey line.

Our initial observation is that data for each organization in Figure
\ref{fig:phdist.figphdists001} is highly skewed and are generally well
fit by decaying power laws.
Four of the six institutions are particularly robust with the
exceptions being Mt.~Sinai Hospital
(Fig.~\ref{fig:phdist.figphdists001}A), which deviates from a simple
power law after the first few hundred donors, and ECHO Science Museum,
which shows the effect of providing strong gift categories, leading in
its case to a shelf at \$1000 (Fig.~\ref{fig:phdist.figphdists001}E).
Examples of similar smaller shelves can be seen in the other
distributions at natural values of \$50, \$100, and so on.

We also see that institutions show remarkable consistency across
years.
For example, in the case of the University of Vermont
(Fig.~\ref{fig:phdist.figphdists001}C), we see Zipf's $\alpha$ and its
related $\gamma$ are relatively stable across three decades of 
gift rank, as
well as over a range of 8,000 to 31,000 gifts per year.
We also see that idiosyncratic distributions such as those of Mount
Sinai (break in scaling, Fig.~\ref{fig:phdist.figphdists001}A) and
ECHO Science Museum (\$1000 shelf,
Fig.~\ref{fig:phdist.figphdists001}E) are strongly preserved from year
to year.

As mentioned above, smaller values of $\gamma$ are associated with
more extreme distributions skewed towards very large gifts.  The two
educational institutions possess extreme distributions with $\gamma
\simeq 1.8$--$1.9 < 2$, and their average gift sizes are relatively
large.  By contrast, United Way has a $\gamma \simeq 2.5 > 2$ meaning
its average gift is small but large ones are possible.

To provide an initial summary, these distributions suggest four
notable characteristics of philanthropic gifting:
\begin{enumerate}
\item 
  The distribution of the size of philanthropic gifts received is
  qualitatively described with a power-law relationship.
\item 
  Within a given institution, the gift-size distribution exponent
  $\gamma$ remains nearly constant year-to-year.
\item 
  As indicated by the similar values of $\gamma$ for the two
  higher education institutions, $\gamma$ may be relatively constant
  within a single philanthropic category.
\item 
  The gift-size distribution exponent $\gamma$ varies considerably
  between philanthropic categories.  
\end{enumerate}

Of the number of questions raised by these observations, we will focus
in particular on one: Why does $\gamma$ vary among different
categories of philanthropic institutions?
Our concrete goal will be to model how giving behavior of people at
different income levels influences the gift-size exponent $\gamma$.

As we show in Fig.~\ref{fig:phdist.Wealth-Giving-dist3a}, there is
considerable variation in donor behavior based on income, and this
provides some insight for our next step forward.
The data we use here comes from the Indiana University Center for
Philanthropy 2005 Study on Charitable Giving by Income Group which
found in particular that a person's income level is strongly
informative of the type of institution they prefer to
support~\cite{copii2007a}.  
Donors earning less than \$100,000 per year, for example, give a higher percent of their philanthropic dollars (8.6\%) to combined purpose funds (e.g. United Way) than do those earning more than \$1,000,000 per year and direct only 4\% of their philanthropic dollars toward such charities.
The opposite is true for education, toward which donors with incomes
less than \$100,000 direct only 3\% of their philanthropic dollars,
while people earning more than \$1,000,000 direct 25\% of their giving.
As such, we would expect that educational institutions would have a
much lower $\gamma$ (associated with higher average gift sizes)
than combined purpose funds.  
Our data bears this out, with a $\gamma$ of \MLEUVMonezerogamma\ for
University of Vermont, and \MLEunitedwayonezerogamma\ for United Way of Chittenden
County (2010).

To make some headway with this issue of varying $\gamma$, 
we first need to examine how gift-size distributions differ in more detail.
In Fig.~\ref{fig:phdist.gammacomparison}, we show that the Albert Einstein
School of Medicine and the United Way of Chittenden County have a
similar sized donor base, both with the 2000th donor giving
approximately \$200, yet the largest gifts to Einstein are roughly 30 times those of the United Way.
How do we explain this?
The answer is not as simple as that all donors to Einstein give a
multiple of what they would give the the United Way: this would not
change the slope (i.e., $\alpha$ or $\gamma$) of the Zipf or frequency 
distributions in log-log space.

As a first attempt, we start with the reasonable assumption that
larger donations originate from wealthier donors.
In terms of Zipf distributions, gift sizes will be ranked in the same
order as the donors who give them, according to their wealth.
If we therefore know, for a given time period, the distribution of the
total amount donated by each individual across a population, we can
estimate how much individuals, as a function of their income, must
relatively give to specific charity categories to obtain the specific
distributions (i.e., values of $\gamma$) we observe in
Fig.~\ref{fig:phdist.figphdists001}.

With this motivation, we turn to evidence that describes how income
and gift-giving are related within the United States.
The 2001 IRS tax return data shown in Figure \ref{fig:phdist.income}
compares reported income to reported charitable deductions.
On average, Americans donated 2.9\% of their income, with deductions
for charitable giving appearing nearly proportionate to income.
When plotted in Zipf format and fitted to a power law distribution,
$\gamma$ for charitable giving ($\gamma=2.41$) is slightly smaller
than that for income ($\gamma=2.56$), favoring donating a slightly
higher percentage of income by wealthier individuals.

To move ahead, we now need to be able to compare two arbitrary Zipf
distributions whether they be Zipf distributions of individual wealth
or gift sizes.
For ease of language, consider gifts given to a specific institution
with $ \alpha=\alpha_{\pinstA} = 1/(\gamma_{\pinstA}-1)$, and total
donations made by individuals in a population with
$\alpha=\alpha_{\popB} = 1/(\gamma_{\popB}-1)$.
We want to know how the first ranked (largest) donation to the
institution compares with the first total amount donated by the
population, and so on, down to the last ranked donation.
We derive this relationship by starting with the Zipf distributions:

\begin{equation} 
  S_{\pinstA}(r) \sim r^{-\alpha_{\pinstA}}
  \mbox{ and } 
  S_{\popB}(r) \sim r^{-\alpha_{\popB}}
  \label{eq:phdist.SASB}
\end{equation}
which, by isolating and equating ranks, immediately gives us
\begin{equation} 
S_{\pinstA}(r)^{-1/\alpha_{\pinstA}}
\sim
S_{\popB}(r)^{-1/\alpha_{\popB}}.
  \label{eq:phdist.SASBconnect}
\end{equation}
Using $\alpha = 1/(\gamma-1)$, 
we then have that the size $S_{\pinstA}(r)$  of a gift 
to the institution
is related to the similarly 
ranked total amount donated by an individual $S_{\popB}(r)$ according to
\begin{equation} 
  S_{\pinstA}(r)
  =
  c
  S_{\popB}(r)^{\, (\gamma_{\popB}-1)/(\gamma_{\pinstA}-1)},
\label{eq:phdist.transformation}
\end{equation}
where
$c=
S_{\pinstA}(r_{\ast})
\left[S_{\popB}(r_{\ast})\right]^{-(\gamma_{\popB}-1)/(\gamma_{\pinstA}-1)}
$
with $r_{\ast}$ being any reference ranking.

We determine how much two individuals $i$ and $j$ in our theoretical population relatively give to the institution.
If these individuals have wealth ranks $r_i$ and $r_j$ (which by assumption are their donation ranks as well), then using~\Req{eq:phdist.transformation}, we have
\begin{equation} 
  \frac{S_{\pinstA}(r_j)}
  {S_{\pinstA}(r_i)}
  =
  \left[
    \frac{S_{\popB}(r_j)}
    {S_{\popB}(r_i)}
  \right]^{\, (\gamma_{\popB}-1)/(\gamma_{\pinstA}-1)}.
\label{eq:phdist.transformation2}
\end{equation}
Finally, we can compute a multiplier $M$ which is the ratio of gift sizes to an institution normalized by the total donated by individuals $i$ and $j$:
\begin{equation} 
  M
  =
  \frac{
    \left[
      \frac{S_{\pinstA}(r_j)}
      {S_{\popB}(r_j)}
    \right]
  }
  {
    \left[
      \frac{S_{\pinstA}(r_i)}
      {S_{\popB}(r_i)}
    \right]
  }
  =
  \left[
    \frac{S_{\popB}(r_j)}
    {S_{\popB}(r_i)}
  \right]^{\, (\gamma_{\popB}-\gamma_{\pinstA})/(\gamma_{\pinstA}-1)}.
\label{eq:phdist.transformation3}
\end{equation}

We can now use~\Req{eq:phdist.transformation3} to transform the distribution of personal giving (and relatedly, that of income) into the distributions for giving to various categories of philanthropy.
First, we estimate $\gamma_{\popB}$ using the 2001 IRS charitable deduction data as a reference distribution, giving $\gamma_{\popB} \simeq 2.41$.
Employing~\Req{eq:phdist.transformation3}, we then calculate multipliers for what people of different total donating levels would have to give to achieve the gift-size distribution exponent $\gamma_{\pinstA}$.

Working from our data sets, we show in Fig.~\ref{fig:phdist.transformingratio} multipliers for six types of institutions, using an income of \$25,000 as an arbitrary reference for convenience.

\begin{figure}[tbp!]
  \includegraphics[width=0.495\textwidth]{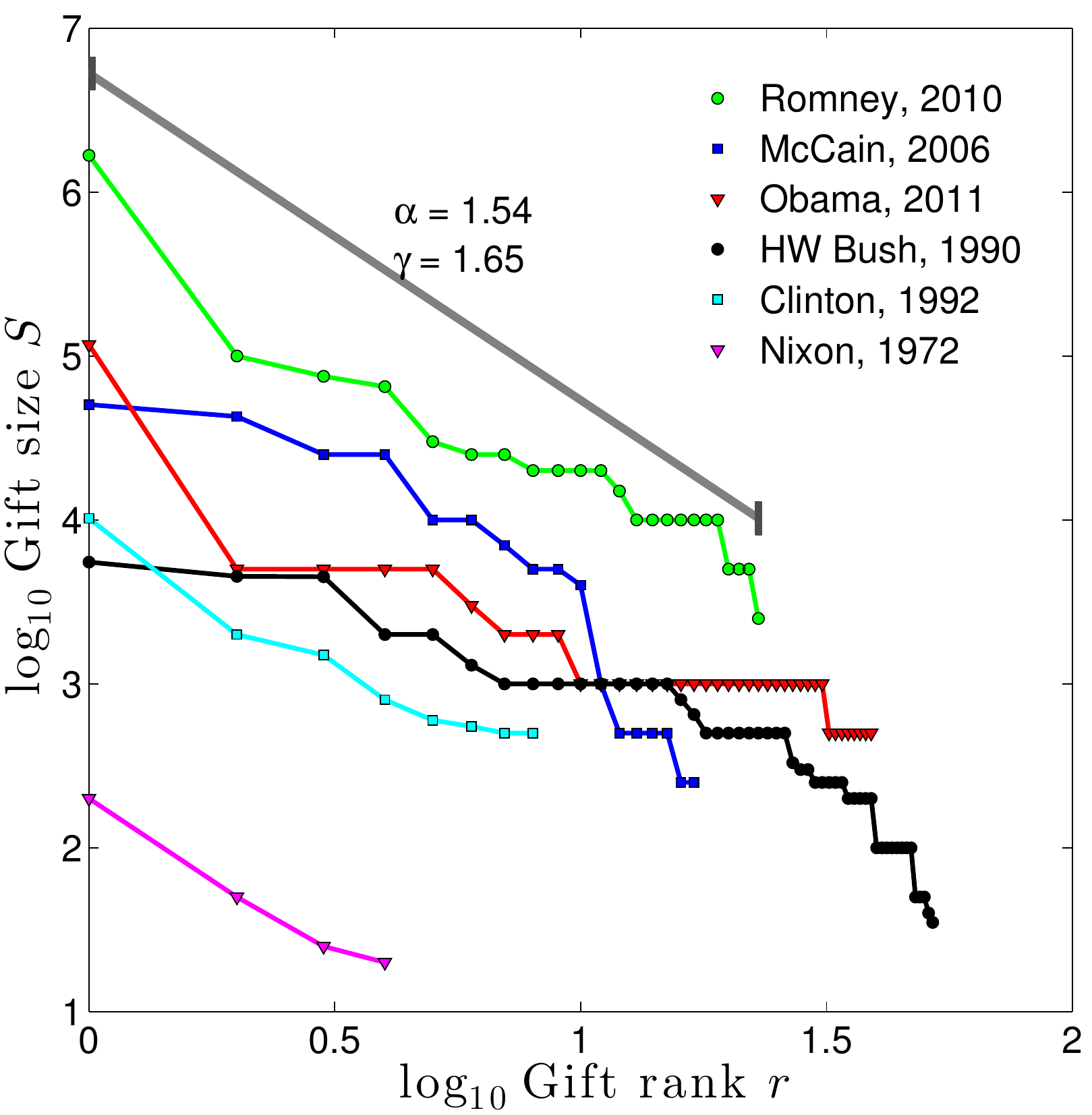}
  \caption{ 
    Charitable gifts of candidates for the United States
    President from their publicly released federal tax returns.
    Again due to finite size bias of maximum likelihood methods,
    we adopted linear regression for fitting the distribution scaling
    parameter $\gamma$.
    The included fit is for President Romney's gifts during the year
    of 2010.
    We include the fitted $\gamma$'s for each president and the range of their
    fit in the supplementary material as Table
    \ref{table:presfitstats},
    and we show comparisons to other distributions in Table
    \ref{table:presfittest}.
  }w
  \label{fig:phdist.presidentialdeductions}
\end{figure}

We see that the multiplier varies strongly across income level and institutional type.
Consider for example that the United Way, which has $\gamma _\text{inst} = \MLEunitedwayonezerogamma$, serves as our example for combined purpose funds.
Because the United Way’s gift-size distribution is fairly close to that of the population’s giving distribution ($\gamma_\text{pop} \simeq 2.41$), the multiplier is close to unity.
Thus, if a person with a total donation level of $S$ directs a certain fraction of their charitable dollars to the United Way, we expect a person with a total donation 10 times as large, $10S$, to also direct a similar fraction of their charitable dollars there as well resulting in a gift approximately 10 times larger (Fig. \ref{fig:phdist.transformingratio}, blue squares).
The multiplier here is 
$10^{(\MLEunitedwayonezerogamma-\MLEEinsteinonezerogamma)/(\MLEEinsteinonezerogamma-1)}
\simeq
0.9,$
which means as a percentage of his total giving, gift from the wealthier person is 0.9 times the gift from the less wealthy person, but in absolute terms, the gift is 9 times larger because his total donation level is 10 times larger.

By contrast, for the University of Vermont for which $\gamma_\text{inst} \simeq \MLEUVMonezerogamma$, the multiplier now depends strongly on income level.
If the same person with a total donation of $S$ were now to direct some of their charitable dollars to the University of Vermont, the higher income person would give a multiplier of 
$ 10^{(\MLEunitedwayonezerogamma-\MLEUVMonezerogamma)/(\MLEUVMonezerogamma-1)}
\simeq
4.0, $
times more of their annual total donations to the same institution (Fig. \ref{fig:phdist.transformingratio}, green circles).
Note that the absolute value of the gift has increased 40 times: 10 times from the larger income, and 4 times from the multiplier effect.

\begin{figure}[tbp!]
  \includegraphics[width=0.495\textwidth]{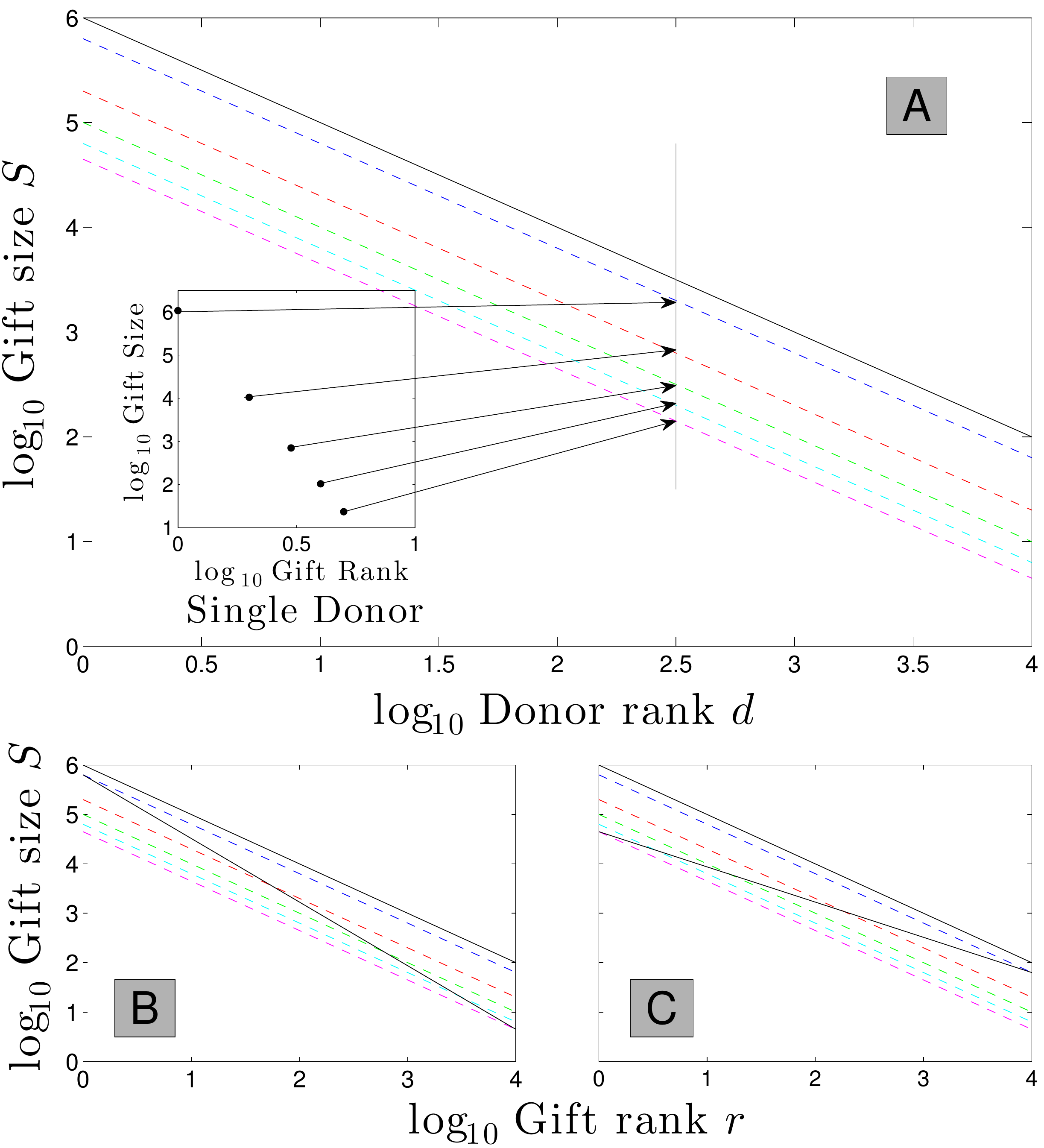}
  \caption{ Model for differing institutional values of $\gamma$.
    \textbf{A.}
    Gift distribution using 2001 IRS deduction $\gamma$ = 2.41 ranks
    the total of each donor's gifts (solid black line), and value of
    donor gifts 1 through 5 (dotted lines).
    Inset shows the five gifts made by donor of rank \#316
    (2.5=$\log_{10}$316) according to a donor $\gamma$ = 1.8. 
    \textbf{B.}
    Institution gift distribution attracting top gift from donor 1,
    and 5th gift from donor 10,000, $\gamma$ = 2.08
    per~\Req{eq:phdistmechanism}.
    \textbf{C.} Institution gift distribution attracting 5th gift from
    donor 1, and top gift from donor 10,000, $\gamma$ = 3.04
    per~\Req{eq:phdistmechanism}.}
  \label{fig:phdist.mechnism001}
\end{figure}

\begin{figure}[tbp!]
    \includegraphics[width=0.495\textwidth]{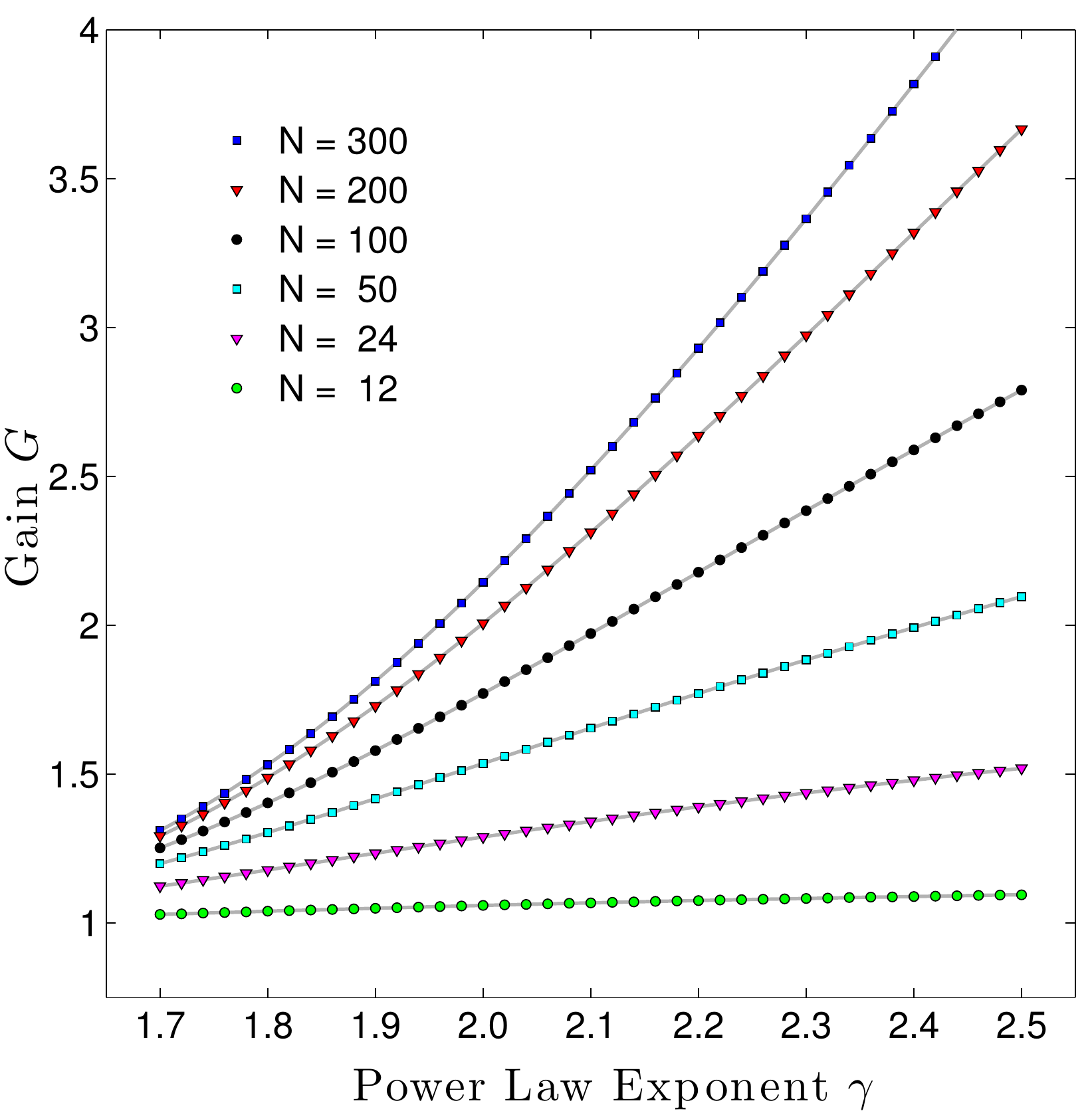}
    \caption{ 
      Expected total amount raised in comparison to amount
      raised by top 12 donors.
      For low $\gamma$ institutions, a larger number of donors has a
      relatively small effect on the total raised.
      For higher $\gamma$ institutions, a large donor pool has a
      greater effect on the total raised.
    }
  \label{fig:phdist.top12_multiplier}
\end{figure}

We can superimpose these multipliers onto the 2005 Study on Charitable Giving by Income Group data.
We do so in Figure~\ref{fig:phdist.Wealth-Giving-dist3b}  which is a rearrangement of the same data displayed in Figure~\ref{fig:phdist.Wealth-Giving-dist3a}, binned according to the IRS income data in Figure~\ref{fig:phdist.income}.
The columns in Figure~\ref{fig:phdist.Wealth-Giving-dist3b} represent data collected from surveys of individual donors.
The lines are the calculated multipliers, shown in Figure ~\ref{fig:phdist.transformingratio}, generated from the values of $\gamma$ shown in Figures ~\ref{fig:phdist.figphdists001} and ~\ref{fig:phdist.OZstory}C.
These are two independent data sets: the former represents data from the gift givers, and the latter is represents data from the gift receivers. 
They agree qualitatively, describing the same story but from different perspectives.

\section{A proposed mechanism for varying slopes}
\label{sec:phdist.mechanism}

We have so far been able to describe how giving patterns must
vary across institutional type as a function of 
institutional giving profiles and donor wealth.
We now attempt to explain in part the origin of these 
variable donation patterns.

In many systems where multiple, dependent power-law size distributions appear,
the exponents involved are typically 
related through simple algebraic expressions~\cite{goldenfeld1992a}.
And while exponents may be tun-able as a function
of independent model parameters~\cite{simon1955a},
smoothly varying relationships between scaling exponents---the kind we have here---are unusual.
Thus we seek to explain part of the giving 
mechanism 
that connects $\gamma_{\popB}$ to $\gamma_{\pinstA}$ 
as being something more than merely ``Power-Law In, Power-Law
Out'' (PLIPLO).

\begin{figure*}[tbp!]
  \includegraphics[width=0.990\textwidth]{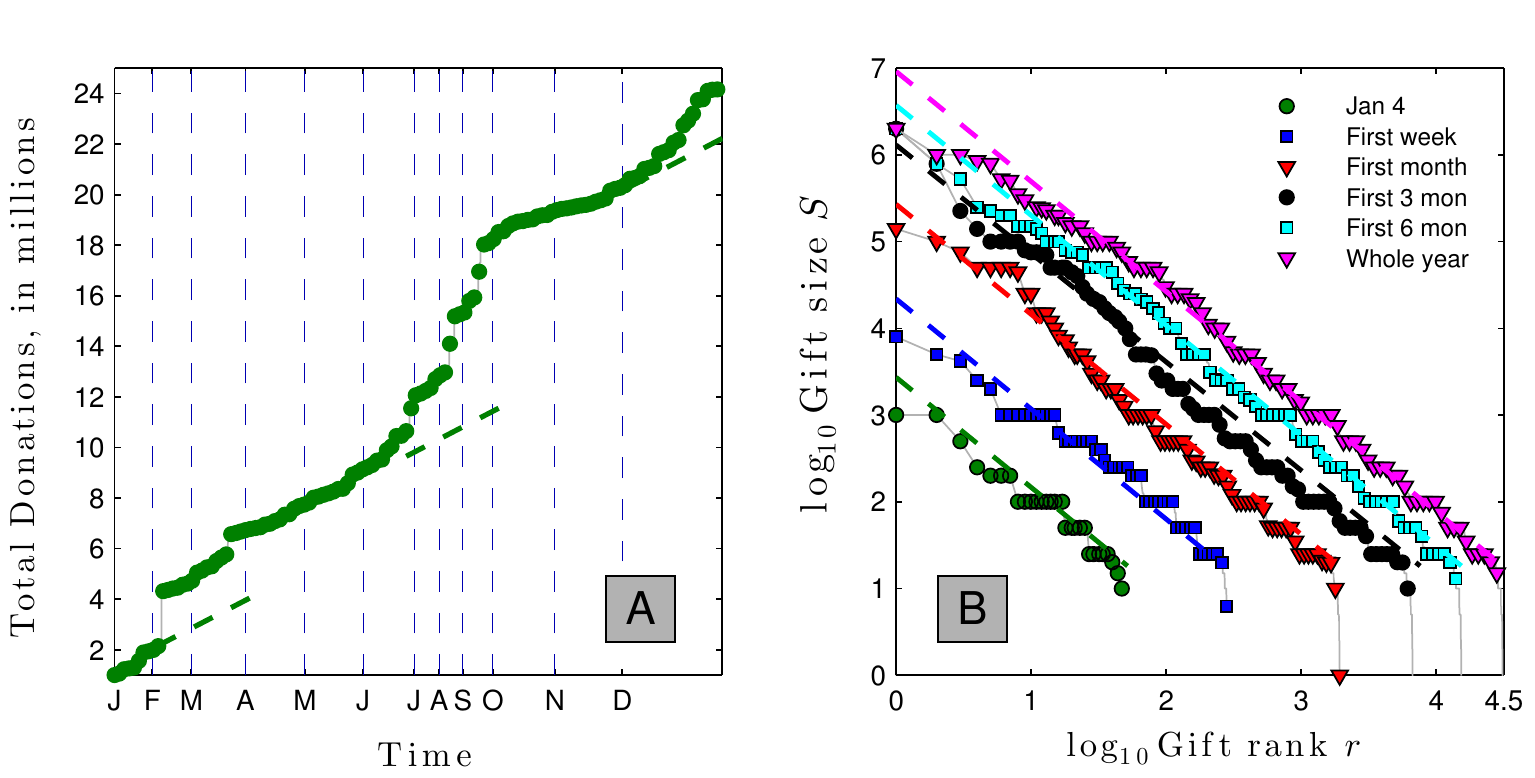}
  \caption{ 
    Panel A: Accumulation of gifts to University of Vermont in
    2010 in the order they were received demonstrates super-linear
    growth.
    Accumulation appears linear until an uncommonly large gift is
    received.
    Panel B: Trend lines intersect the vertical axis at the expected maximum
    gift and show how the expected maximum gift grows with the number of
    donors according 
    to~\Req{eq:phdistmechanism}.
  }
  \label{fig:phdist.uvmaccumulation}
\end{figure*}

\begin{figure*}[tbp!]
  \includegraphics[width=\textwidth]{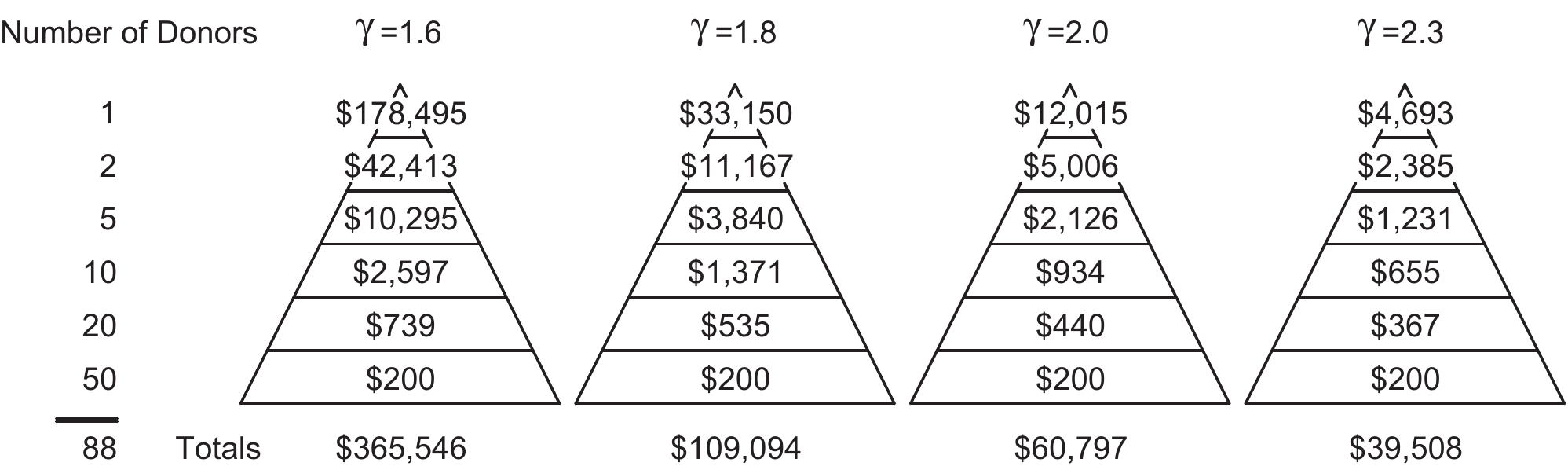}
  \caption{ Fundraising pyramids customized to an institutions $\gamma$.
    Using a power-law model of fundraising, low $\gamma$ institutions
    should plan for and request much higher top level gifts than high
    $\gamma$ institutions.
  }
  \label{fig:phdist.pyramid}
\end{figure*}

We start by looking at the giving behavior of individual donors, who
will differ in terms of the number of gifts they make, the size of
these gifts, and their personal ranking of target institutions.
They may choose to make their largest gift to health, their next
largest to education, and so on.
We would like examine data that characterizes the giving behaviors of
these donors, such as through examination of itemized charitable
deductions on federal tax returns.
While this private information is generally inaccessible, some
presidential candidates have released their tax returns publicly, and
we can use this data as a rough guide.
Fig.~\ref{fig:phdist.presidentialdeductions} shows itemized deductions
for several candidates, plotted on log-log scales.
These donations, ranked largest to smallest, are visually consistent
with an approximate power law Zipf distribution with gamma ranging
from 2 to 3, with an average around 2.5 
(see the supplementary material, 
Tabs.~\ref{table:presfitstats} and \ref{table:presfittest}).
For simplicity, we will again presume that gifts made by
individual donors can be adequately described by a power law Zipf
distribution.
We can then propose a mechanism that uses donor choices to explain the
different gammas we see among philanthropic institutions.

In Fig.~\ref{fig:phdist.mechnism001}A, we assume a gift size
distribution with $\gamma_{\popB} = 2.41$ and plot the top five gifts
using a rough estimate of $\alpha_{\rm donor} = 1.8$.
The inset shows the head of the Zipf distribution for an example
donor.

Fig.~\ref{fig:phdist.mechnism001}B shows the gift distribution for an
institution with strong appeal to the \#1 donor garnering their top
gift, but interest in this institution monotonically decreases among
ranked donors until it attracts the 5th gift from the final donor.
This generates a low $\gamma$ consistent with, for example, the
Einstein School of Medicine illustrated in
Fig.~\ref{fig:phdist.gammacomparison}.
Fig.~\ref{fig:phdist.mechnism001}C shows the opposite arrangement of
donor appeal, leading to a high $\gamma$ profile more typical of a
religious institution.

Following this prescription we can straightforwardly derive an
institution's Zipf exponent $\alpha_{\pinstA}$ (and corresponding
$\gamma_{\pinstA}$) as a function of a donor's Zipf exponent
$\alpha_{\rm donor}$, the population's Zipf exponent $\alpha_{\rm
  pop}$ (e.g., from the IRS charitable deduction distribution), the
rank of the first donor's and final donor's gift choice, and the
number of donors $N$, As defined by our power law model, such a
relationship is linear in log space, the slope of which equals
$-\alpha_{\pinstA}$.
We have
\begin{align*}
\alpha_{\pinstA} = -\frac{\left ( \begin{aligned} &\log_{10}(\text{Last ranked donor's  gift})\\ &~~~-\log_{10}(\text{First ranked donor's gift}) \end{aligned} \right )}{\log_{10} N- \log_{10} 1} \end{align*} 
where the size of the first ranked donor's gift is given by
\begin{equation*}
k  (\text{Rank of first donor's choice})^{-\alpha_{\text{donor}}}
\end{equation*}
and the size of the last ranked ($N$th) donor's gift is
\begin{equation*}
k (\text{Rank of final donor's choice})^{-\alpha_{\text{donor}}} N^{-\alpha_{\popB}}.
\end{equation*}
Substituting these equations into the equation for slope and
simplifying gives the relationship we seek:
\begin{equation} \label{eq:phdistmechanism}
\alpha_{\pinstA}
= 
\alpha_{\popB}  
+  
\frac{\alpha_\text{donor}}{\log_{10} N}
\log_{10} 
\left(
  \frac{{\rm Rank\ of\ final\ donor's\ choice}}
  {{\rm Rank\ of\ first\ donor's\ choice}}
\right).
\end{equation}
For the example Fig.~\ref{fig:phdist.mechnism001}B and C,
the above gives a range of $\gamma_{\pinstA}$ from around 2 to 3.
While the rank of first and final donors' choices are fixed integers,
the relationship will in practice be statistical.

\section{Recommendations for Fundraisers}
\label{sec:phdist.recommendations}

\subsection{The Top-12 Rule}
\label{subsec:phdist.top12rule}

When undertaking a capital campaign, an institution will want to
estimate the fundraising capacity of its community.
Capital campaigns tend be be a more focused fundraising effort
targeting fewer donors than the annual campaigns reported in this
paper.
We have collected some preliminary data suggesting an institution's
capital campaign $\gamma$ tends to be less than that of its annual
campaign, resulting in a more extreme distribution of gifts.
Dove reports that the top 10 to 15 donors commonly account for 50 to
70 percent of total funds raised~\cite{dove2000a}.
Similarly, a professional consultant estimates capital
campaign fundraising capacity using a rule-of-thumb that the top 12
donors will contribute 65\% of the revenue~\cite{graham2011b}.
For the sake of discussion, let us refer to this as the top-12-rule.
If we can estimate an expected $\gamma$ for the campaign, have an idea
of the expected number of donors N, and know how much to expect from
the 12 largest gifts, we can calculate a gain factor $G$ that will give
an estimate for the campaign total:
\begin{equation}
  \text{Campaign Total} = G \times \sum_{r=1}^{12} S(r))
\end{equation}
where
\begin{equation} 
  G 
  =
  \frac{\sum_{r=1}^{N} S(r)}
  {\sum_{r=1}^{12} S(r)}
          \end{equation}

When we apply a power-law model of philanthropic giving to the
top-12-raises-65\% rule, we can find circumstances for when this rule
applies, and when it does not.
Fig.~\ref{fig:phdist.top12_multiplier} confirms that for campaigns
with values of $\gamma$ in the 1.8 to 1.9 range (e.g., for higher education),
the total raised is about 1.5 times that of the top 12 donors, and
increases only marginally for a donor pool total of 300 versus 100.
But the rule grossly underestimates the total raised for institutions
with larger values of $\gamma$.
For a campaign with a $\gamma$ around 2.3 (e.g., for combined
purpose funds) we expect the total raised by 100 donors to exceed
twice that of its top 12, and that 300 donors would triple the total
of its top 12.
This analysis would suggest that the top-12-raises-65\% rule is a poor
fundraising estimator for higher values of $\gamma$ typical of combined purpose
funds and religious organizations, especially for campaigns with a
high number of expected donors.

Note that Fig.~\ref{fig:phdist.top12_multiplier} is predicated on
the assumptions that we can identify the top 12 donors for a group of
a given size and that a power-law distribution of gifts applies
throughout that donor pool.
These assumptions no longer apply if we then increase the size of the
original donor pool, because gifts from the new donors will not add
serially to the tail of the distribution, but will populate all
positions throughout the distribution and may exceed some of the
original top 12 gifts as the pool is enlarged.
This would have the effect of raising somewhat more money than
predicted by Fig.~\ref{fig:phdist.top12_multiplier}.

In other words, as a campaign extends its original scope it may
receive a gift within or greater than the original top 12.
For a power-law model of growth, the total amount raised tends to grow
super-linearly with the total number of donors: the {\em expected}
amount raised from 100 donors is more than twice the {\em expected}
amount raised from 50 donors (provided the expanded pool of donors has
the same characteristic wealth distribution and interest in the
organization as the original pool).
This trend, however, is subject to great variability, and is more
pronounced for smaller than larger values of $\gamma$.
The expected largest gift from this enlarged group of donors follows
the scaling:
\begin{equation}
  \frac{\text{Max gift in group } A}
  {\text{Max gift in group } B}
  =
  \left(
    \frac{\text{\# donors in group } A}
    {\text{\# donors in group } B}
  \right)^{1/(\gamma-1).}
  \label{eq:phdistphil.maxgift}
\end{equation}

The following example demonstrates the haphazard variability that this relationship is subject to.
Fig.~\ref{fig:phdist.uvmaccumulation}A shows the accumulating total in the order that gifts were received at University of Vermont in 2010.  
The total appears to grow linearly for a while, then jumps upwards when an exceptionally large gift is received.

Fig.~\ref{fig:phdist.uvmaccumulation}B shows these gifts broken down into various time frames.
The dotted lines demonstrate how the {\em  expected} largest gift grows (per~\Req{eq:phdistmechanism}) as the number of donors increases.
Not surprisingly, the {\em actual} largest gift shows substantial
variability around this predicted value.

\begin{figure}[tbp!]
  \includegraphics[width=0.475\textwidth]{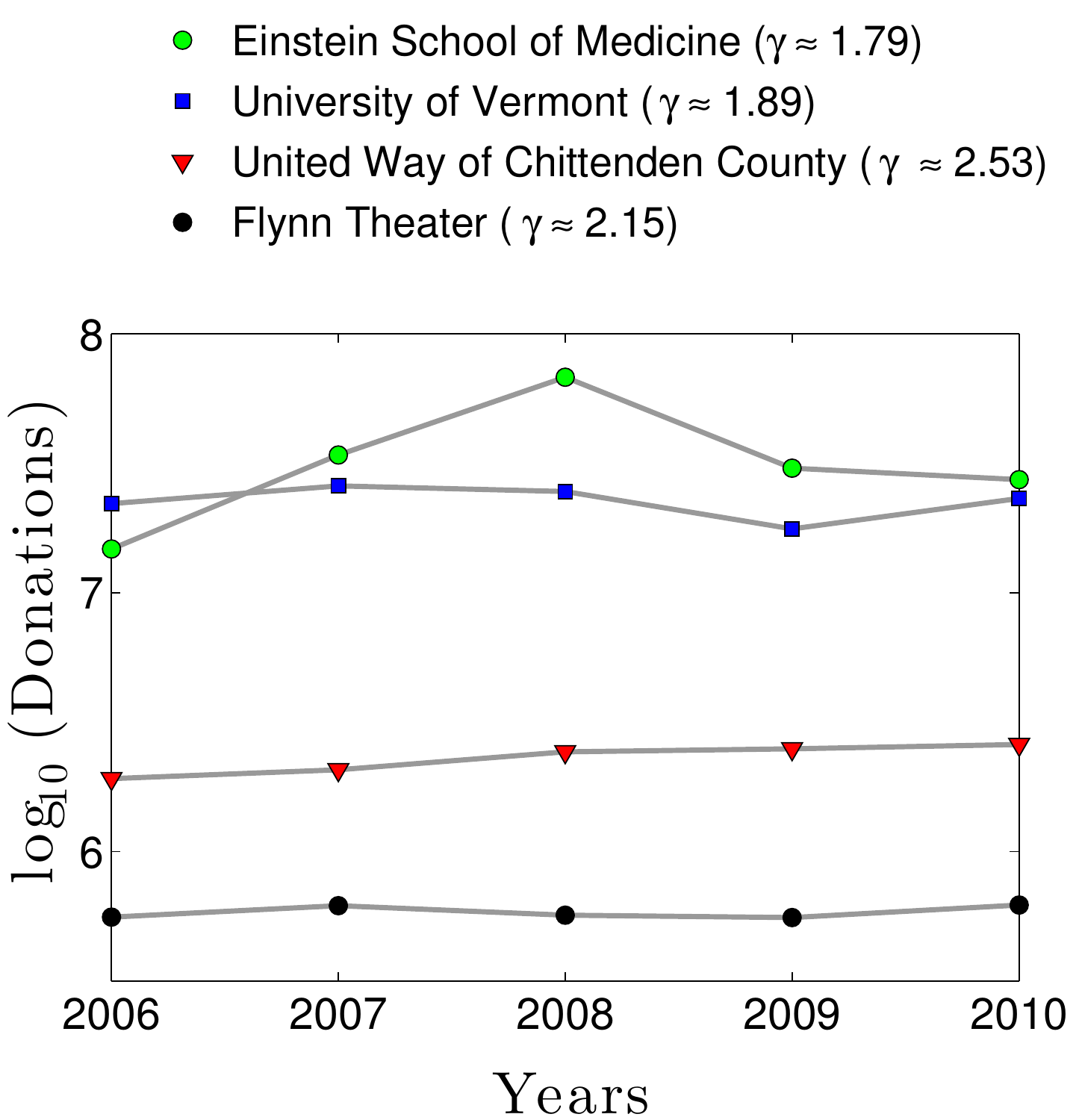}
  \caption{
    Low $\gamma$ organizations can expect to have greater year-to-year fundraising volatility than higher $\gamma$ organizations.
    The $\gamma$ values reported here are averages for all of the years shown in Figure \ref{fig:phdist.figphdists001}.
  }
  \label{fig:phdist.variability}
\end{figure}

\begin{figure*}[tbp!]
  \centering
  \includegraphics[width=\textwidth]{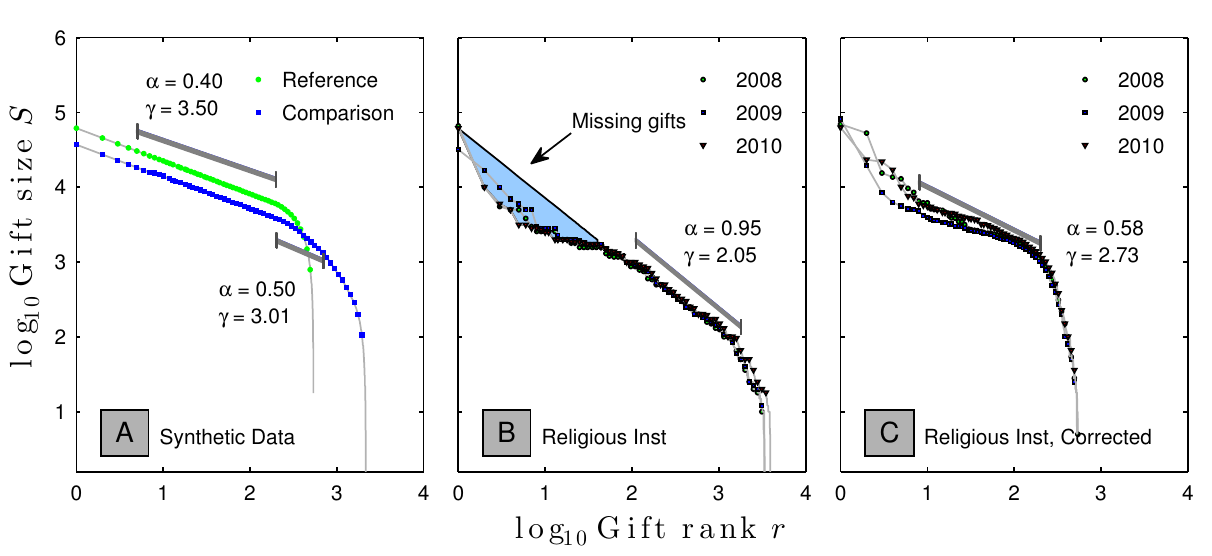}
  \caption{ 
    \textbf{A.}
    `Reference'' synthetic power law giving
    distribution with a largest gift of \$62,000, $N$=542 total donors
    of which the largest 200 follow a power law with $\gamma = 3.50$
    is created and labelled Reference, with a fit show between the 5th
    and 200th gifts.
    A comparison is created where each gift from the reference
    distribution is split into 60\%, 25\%, 10\%, and 5\%.
    A false shoulder is created with a $\gamma = 3.01$, fitting the
    slope between the 200-th and 700-th donors.
    \textbf{B.}
    Gift-size data from an anonymous religious institution.
    The blue region appears to represent around \$100,000 of
    unrealized potential in the \$1,000 and above range.
    The power law slope of $\gamma = 2.04$ is found by fitting the gifts from donor
    110 to donor 1800, over the region that appears flattest.
    \textbf{C.}
    The giving data from panel \textbf{B} corrected for multiple
    donations, and a correct $\gamma$ of 2.73 is found between the 8th
    and 200-th donors.
  }
  \label{fig:phdist.OZstory}
\end{figure*}

\subsection{The 80-20 Principle and the Fundraising Pyramid}
\label{subsec:phdist.pyramid}

Commonly the gifts table (fundraising pyramid) is predicated on
Vilfredo Pareto's 80/20 principle; that 80\% of funds are raised from
20\% of donors~\cite{pierpoint1998a}.
Pareto originally founded his principle on his observation of the
power-law size distribution of wealth in Italy~\cite{newman2005b} (the
$\gamma$ for an 80/20 fundraising relationship varies based on the
number of donors, ranging from 1.82 for 100 donors, to 2.04 for 5,000
donors).
By knowing an organization's $\gamma$ and donor pool size, a
fundraiser can develop a giving pyramid specific to that
organization's gift distribution, rather than using a generic 80/20
pyramid.
Fig.~\ref{fig:phdist.pyramid} shows four pyramids, each calculated
for an organization with a specific $\gamma$.
For each pyramid the lowest level of donations is set at \$200.
Organizations with lower values of $\gamma$ should plan for much larger gifts at
the higher levels than organizations with higher values of $\gamma$, and should
expect to raise much more from the same number of donors.

\subsection{What $\gamma$ says about an organization's fundraising capacity and robustness}
\label{subsec:phdist.capacity}

Low values of $\gamma$ generally describe organizations whose largest gifts are
extremely large in proportion to the total amount raised.
For example, for a campaign with a $\gamma$ of \MLEUVMonezerogamma\
(e.g., higher education) and 1000 donations, the lead gift would be
expected to be about 30\% of the total raised.
But for a campaign with a $\gamma$ of \MLEunitedwayonezerogamma\ (e.g., combined
purpose fund) the lead gift would be expected to be about 5\% of the
total.
In contrast to the predictability of mid-level gifts, lead gifts are
subject to enormous variability about their expected value, as can be
seen by their divergence from the projected line of slope in most of
the data examples presented in this paper.
This means that while low $\gamma$ institutions are more likely to
enjoy the benefit of extremely large gifts, their annual fundraising
total is highly dependent on the gifts of those top few donors and
becomes subject to significant year-to-year variability
(Fig.~\ref{fig:phdist.variability}).
In contrast, high $\gamma$ institutions are likely to experience more
stable year-to-year totals.
Note that United Way of Chittenden County and Einstein School of
Medicine have a similar sized donor base, demonstrating the
fundraising power of a low $\gamma$.  

\subsection{Misleading Effect of Multiple Donations per Donor}
\label{subsec:phdist.datacollection}
For proper analysis, data for a given time period must reflect a
single total of gifts from each donor.
If the donor has made multiple gifts, and these gifts are recorded
separately, the number of donors, $N$, will be falsely inflated.
This can create a false and misleading shoulder on the institution's
Zipf plot and lead to a miscalculation of $\gamma$.
Panel B of Fig.~\ref{fig:phdist.OZstory} appears to show gifts from
3,500 donors to a(n anonymous) religious institution, with $\gamma$ of
2.04.
There appears to be about \$100,000 of unrealized potential from the
larger gifts.
In fact, there were only 500 donors, but many of them had made
multiple smaller gifts throughout the year.
In panel C, for each year we properly summed multiple gifts by single
donors into a single total for each donor, and see that $\gamma$ of
2.04 in Panel B was entirely due to a false shoulder effect.
The correctly measured $\gamma$ of 2.73 is more consistent with that
predicted from the donor survey data for religious institutions from
Fig.~\ref{fig:phdist.Wealth-Giving-dist3b}.
We now see that the largest gifts in fact exceeded expectations.

\section{Concluding Remarks:}
\label{sec:phdist.phdist.conclusion}
The distribution of gifts received by nonprofit institutions is
approximately consistent with a power-law size model.
Individual institutions, and possibly broad of categories
institutions, have their own characteristic scaling exponent $\gamma$.
Fundraising projections modeled on power laws may be useful for
predicting the success of a given campaign, 
and for affecting the
strategic planning of a campaign.

Future study should assemble a larger database to see if our findings
are consistent, to study if there is a predictable relationship
between the values of $\gamma$ for an institution's annual fund and its capital
campaigns, and to capture values of $\gamma$ for new philanthropic categories such
as human services and the environment.
Different regions across the globe and within countries may show characteristic local
variations for values of $\gamma$ for income, overall giving, and by category of
institution.
Gifts from private individuals account for 73\% of giving; family
foundations, corporate giving, and bequests account for the
remainder~\cite{givingusa2011a}.
Analysis of these disparate funding sources may find characteristic
values of $\gamma$ for gifts based on the category of their source in addition to the category of their destination.

\acknowledgments

We would like to extend thanks to the development staff at 
the Albert Einstein College of Medicine, 
Mount Sinai Hospital, 
the University of Vermont, 
United Way of Chittenden County, 
the Flynn Theater, 
and 
the ECHO Lake Aquarium and Science Center, 
without whose help this research would not be possible;  
and to Christine Graham for contributing her experience as a professional fundraiser.
\revtexonly{PSD was supported by NSF CAREER Award \# 0846668. The authors are grateful for the computational resources provided by the Vermont Advanced Computing Core which is supported by NASA (NNX 08A096G)}

\clearpage

\appendix

\setcounter{page}{1}
\renewcommand{\thepage}{S\arabic{page}}
\renewcommand{\thefigure}{S\arabic{figure}}
\renewcommand{\thetable}{S\arabic{table}}
\setcounter{figure}{0}
\setcounter{table}{0}

\section*{Supplementary material}
\label{sec:phdist.supp}

All data sets can be downloaded from our present paper's online appendix
which is located here: \url{http://www.uvm.edu/storylab/share/papers/gottesman2014a/}.

In this supplementary section, we provide some minor details 
for the data used in the main paper.
We also show results of fitting gift size distributions
to a number of potential candidate forms,
employing methods of Maximum
Likelihood (ML) for the estimation of parameters~\cite{clauset2009b}.
Unsurprisingly, a pure power law decay does not fit
the data with great precision.
Nevertheless, we justify our use of a power law
size distribution as a reasonable, if rough, characterization 
of philanthropic gift size distributions---very much in the
manner of standard linear regression---and hence a suitable building block for our analyses.
Larger, much more exhaustive data sets across all kinds
of institutions will be required to strongly advance our knowledge
of philanthropy beyond what we have achieved here.

\subsection{Further details concerning philanthropic data}

For all sources of data in this present work, we made no distinctions as to whether the
donor of a gift was a living person, a bequest, a foundation, or a
corporation.
In the 5 year period 2006 through 2010, the sources of total giving in
dollars were divided as 73\% individuals, 8\% bequests, 14\% foundations,
and 5\% corporate~\cite{givingusa2011a}.

For the institutions analysed in the main paper, 
gift size specifics were as follows:
\begin{itemize}
\item 
  Albert Einstein School of Medicine, University of Vermont, ECHO
  Aquarium and Science Center, and the Flynn Theater provided data for
  all gifts received over 5 years.
  Multiple gifts by a single donor over a single year were not
  identified by these institutions, and hence not summed into a single gift.
\item
  Mount Sinai Hospital reported all gifts and was able to identify
  multiple gifts per year from individual donors; these were summed into
  a single gift per donor per year.
\item
  United Way of Chittenden County likewise identified multiple gifts
  which were summed into a single gift per donor per year, but was able
  to provide data only on gifts received individually.
  Some workplaces collect United Way donations and then send a lump sum:
  these sums do not reflect individual gifts and were not reported to
  us.
\item
  The anonymous religious institution described their gifts both as
  multiple donations per donor per year and a summed donation per donor
  per year, permitting construction of Fig.~\ref{fig:phdist.OZstory}.
\end{itemize}

Individual donations from United States Presidents and candidates were
obtained directly from their tax returns for the stated years.
This data is available directly at 
\url{http://www.taxhistory.org/www/website.nsf/Web/PresidentialTaxReturns}.
In addition, we include the data presented here in a CSV file.

\subsection{Scaling parameter fitting}

\begin{figure*}[h!]
\begin{center}
  \includegraphics[width=0.695\textwidth]{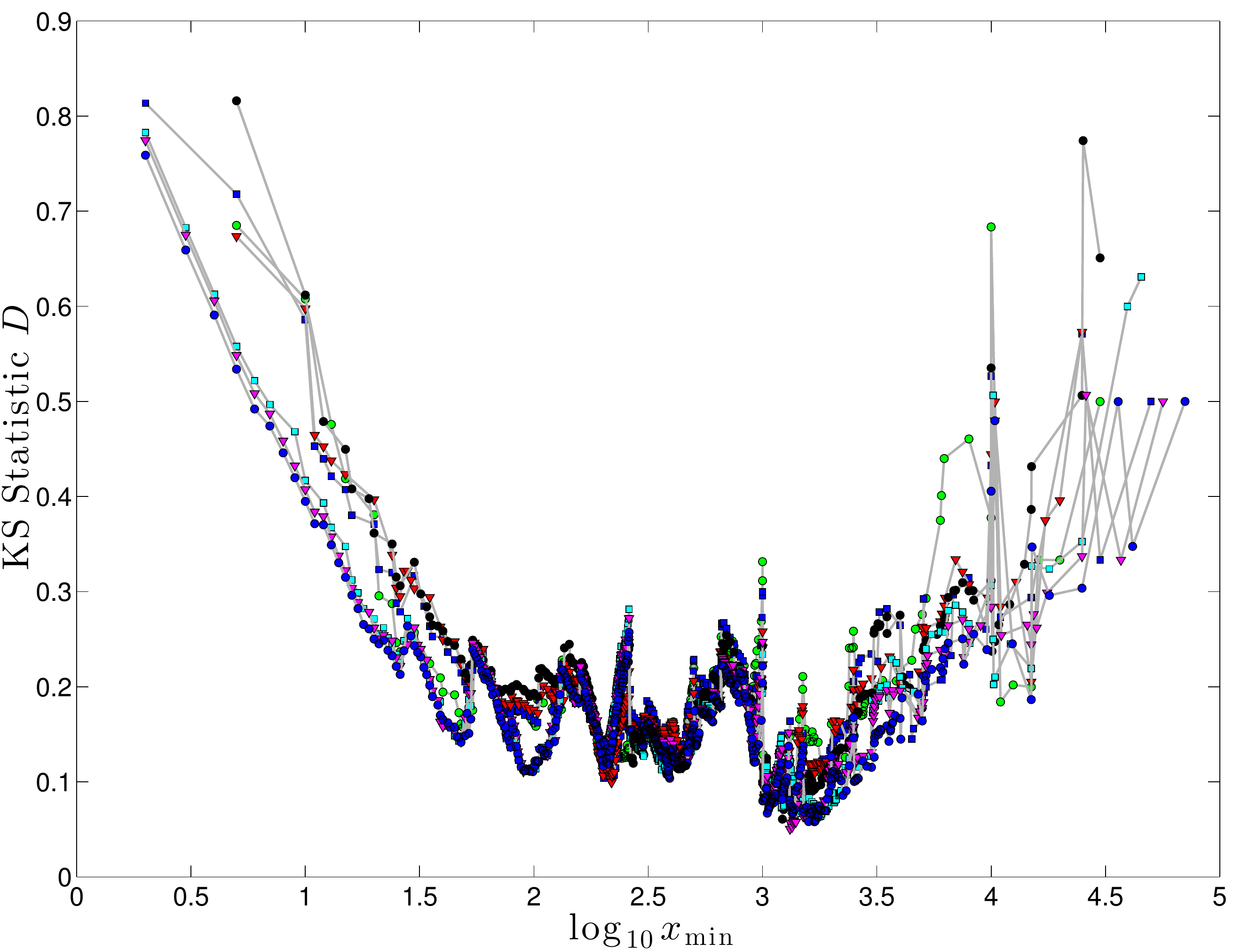}
  \caption{ 
    The Kolmogorov-Smirnoff statistic $D$ plotted over $\log_{10}
    x_{\min}$, where $x_{\min}$ is the minimum value fit for power law
    behavior, for the United Way of Chittenden County over the years
    2006--2010.
    $D$ is generated from the ML estimate.
    Existence of multiple minima in our data indicate that there are
    multiple possible fitting regions for which the KS statistic
    suggests a good fit.
    The variability of this value over each year plotted produced
    widely varying scaling parameters $\gamma$, and thus could not be
    safely used.
  }
  \label{fig:KSvsD}
\end{center}
\end{figure*}

In general, we utilize the ML method to fit our scaling parameter
$\gamma$.
However, for small data (presidential gifts and limited tax data) the
ML method is biased from the finite size and we use a linear
regression for a rough estimate.
To determine the portion of our data that is best power-law behaved, the
minimization of the Kolmogorov-Smirnoff statistic $D$ proved to be
inconsistent across our data due the multiple minima of the statistic
$D$ (Figure \ref{fig:KSvsD}).
For this reason, we empirically chose the scaling regions (i.e., the cut offs).

In Tab.~\ref{table:fitstats},
we report results for fitting power-law decay distributions using the
ML approach.
(In Tab.~\ref{table:presfitstats} we perform the same analysis for presidential gifts.)
Some code from both Clauset and Alstott was used, in addition to our
methods~\cite{clauset2009b,alstott}.
We note that as argued in Alstott, the $p$-value of the fitted
distribution becomes less useful for large data sets because the Monte
Carlo generated distributions become nearly perfect~\cite{alstott}.
Since our data is large, we find that in general none of the
synthetically generated data sets has $D$ greater than for the real
data ($p = 0.00$), but this does not rule out power law behavior.

We then turn to the comparison to other distributions in
Tab.~\ref{table:fittest} 
(and Tab.~\ref{table:presfittest} for presidential gifts), and find that the power law is at
least reasonably supported in most cases.
The only test for which statistical significance was at least common
is for the exponential distribution, and in all cases the power law was favored.
For all distributions where the cutoff power law test was significant, in
particular the Einstein School of Medicine, we find that the cutoff power
law is favored, and would be the most likely distribution.

\begin{table*}
\begin{tabular}{|lc|rrr|cl|cc|}
\hline
{\bf Institution} & {\bf Year} & $~~~~~~~~~~~\mathbf{ \langle x \rangle}$ & $~~~~~~~~~~~~~\mathbf{ \sigma}$ & $~~~~~~~~~\mathbf{x_\text{max}}$ &  $~~~~~~~~~\mathbf{\gamma}~~~~~~~~~$ & $\mathbf{\text{Range} }~~~~~$ & ~~{\bf D}~~ & ~~~{\bf p} ~~~\\ 
\hline
Mount Sinai Hospital & 2009 & 17618.40 & 450408.65 & 37259947 & 1.92 $\pm$ 0.08 & 1 to 90 & 0.12 & 0.00\\
 & 2010 & 19348.18 & 429587.88 & 27885708 & 2.02 $\pm$ 0.10 & 1 to 90 & 0.10 & 0.00\\
\hline
Einstein School of Medicine & 2006 & 3247.30 & 46940.29 & 2000000 & 1.79 $\pm$ 0.02 & 1 to 2000 & 0.11 & 0.00\\
 & 2007 & 4768.09 & 78762.48 & 5350000 & 1.71 $\pm$ 0.01 & 1 to 2000 & 0.15 & 0.00\\
 & 2008 & 10385.80 & 199751.68 & 10200000 & 1.80 $\pm$ 0.01 & 1 to 2000 & 0.21 & 0.00\\
 & 2009 & 5212.92 & 139468.89 & 10000000 & 1.84 $\pm$ 0.01 & 1 to 2000 & 0.15 & 0.00\\
 & 2010 & 4917.94 & 61893.49 & 2000000 & 1.80 $\pm$ 0.06 & 1 to 2000 & 0.15 & 0.00\\
\hline
Univeristy of Vermont & 1974 & 155.76 & 2811.94 & 200000 & 1.94 $\pm$ 0.01 & 3 to 794 & 0.18 & 0.00\\
 & 1980 & 284.31 & 5284.36 & 326000 & 1.85 $\pm$ 0.03 & 3 to 794 & 0.11 & 0.00\\
 & 1990 & 350.23 & 5382.45 & 500000 & 2.16 $\pm$ 0.01 & 3 to 794 & 0.38 & 0.00\\
 & 2000 & 805.33 & 15120.53 & 1488000 & 1.71 $\pm$ 0.03 & 3 to 794 & 0.09 & 0.00\\
 & 2010 & 741.40 & 17029.10 & 2000000 & 1.81 $\pm$ 0.05 & 3 to 794 & 0.13 & 0.00\\
\hline
United Way, Chittendon County & 2004 & 441.71 & 1133.02 & 30000 & 2.77 $\pm$ 0.04 & 1 to 316 & 0.21 & 0.00\\
 & 2005 & 464.47 & 1444.26 & 50000 & 2.58 $\pm$ 0.22 & 1 to 316 & 0.13 & 0.00\\
 & 2006 & 456.86 & 1199.92 & 25000 & 2.42 $\pm$ 0.05 & 1 to 316 & 0.07 & 0.00\\
 & 2007 & 456.16 & 1279.14 & 30000 & 2.42 $\pm$ 0.14 & 1 to 316 & 0.07 & 0.00\\
 & 2008 & 287.53 & 1089.92 & 45460 & 2.53 $\pm$ 0.00 & 1 to 316 & 0.14 & 0.00\\
 & 2009 & 278.93 & 1122.44 & 56500 & 2.55 $\pm$ 0.08 & 1 to 316 & 0.12 & 0.00\\
 & 2010 & 287.58 & 1271.10 & 70518 & 2.47 $\pm$ 0.09 & 1 to 316 & 0.08 & 0.00\\
\hline
ECHO Science Museum & 2005 & 977.77 & 3153.41 & 25000 & 1.66 $\pm$ 0.03 & 2 to 88 & 0.20 & 0.00\\
 & 2006 & 951.16 & 3415.22 & 25000 & 1.59 $\pm$ 0.02 & 2 to 88 & 0.28 & 0.00\\
 & 2007 & 941.61 & 3161.08 & 25000 & 1.59 $\pm$ 0.07 & 2 to 88 & 0.31 & 0.00\\
 & 2008 & 956.88 & 2688.31 & 20000 & 1.56 $\pm$ 0.01 & 2 to 88 & 0.26 & 0.00\\
 & 2009 & 676.84 & 2098.96 & 20000 & 1.73 $\pm$ 0.15 & 2 to 88 & 0.17 & 0.00\\
\hline
Flynn Theater & 2006 & 241.87 & 1528.82 & 65065 & 2.18 $\pm$ 0.04 & 1 to 2000 & 0.26 & 0.00\\
 & 2007 & 268.54 & 1732.33 & 60000 & 2.15 $\pm$ 0.05 & 1 to 2000 & 0.25 & 0.00\\
 & 2008 & 248.00 & 1015.39 & 27500 & 2.15 $\pm$ 0.00 & 1 to 2000 & 0.22 & 0.00\\
 & 2009 & 242.90 & 1212.42 & 40000 & 2.18 $\pm$ 0.04 & 1 to 2000 & 0.23 & 0.00\\
 & 2010 & 246.13 & 1606.43 & 70000 & 2.09 $\pm$ 0.05 & 1 to 2000 & 0.22 & 0.00\\
\hline
\end{tabular}
\caption{Summary statistics of all of the donation data is presented. The reported $\gamma$ and range are fit with the MLE method, and the $x_\text{min}$ which was found to minimize the Kolmogorov-Smirnoff statistc {\bf D} is reported along with {\bf D} itself. In this case, lower values of {\bf D} indicate a better fit.}
\label{table:fitstats}
\end{table*}
\begin{table*}
\begin{tabular}{|lc|c|cc|cc|cc|cc|}
\hline
 &  & & \multicolumn{2}{c}{{\bf Log-Normal }} & \multicolumn{2}{|c|}{{\bf Exponential }} & \multicolumn{2}{|c|}{{\bf Stretched Exp. }} & \multicolumn{2}{|c|}{{\bf Cutoff Power Law}} \\ 
{\bf Institution} & {\bf Year} & {\bf ~~p~~} & {\bf ~~LR~~} & {\bf ~~p~~}  &  {\bf ~~LR~~} & {\bf ~~p~~} & {\bf ~~LR~~} & {\bf ~~p~~} & {\bf ~~LR~~} & {\bf ~~p~~}\\ 
\hline
Mount Sinai Hospital & 2009 & 0.00& -0.21 & 0.67& \textcolor{blue}{31.80} &  \textcolor{blue}{$\mathbf{0.01}$}& -0.19 & 0.82& -0.53 & 0.30\\
 & 2010 & 0.00& -0.00 & 0.99& \textcolor{blue}{47.31} &  \textcolor{blue}{$\mathbf{0.00}$}& 0.46 & 0.60& -0.23 & 0.50\\
\hline
Einstein School of Medicine & 2006 & 0.00& \textcolor{red}{-6.22} &  \textcolor{red}{$\mathbf{0.03}$}& \textcolor{blue}{378.82} &  \textcolor{blue}{$\mathbf{0.00}$}& \textcolor{red}{-7.06} &  \textcolor{red}{$\mathbf{0.03}$}& \textcolor{red}{-8.31} &  \textcolor{red}{$\mathbf{0.00}$}\\
 & 2007 & 0.00& -0.30 & 0.59& \textcolor{blue}{17.65} &  \textcolor{blue}{$\mathbf{0.01}$}& -0.35 & 0.61& -0.67 & 0.25\\
 & 2008 & 0.00& -1.03 & 0.37& \textcolor{blue}{1235.22} &  \textcolor{blue}{$\mathbf{0.00}$}& 0.71 & 0.81& \textcolor{red}{-2.85} &  \textcolor{red}{$\mathbf{0.02}$}\\
 & 2009 & 0.00& -2.48 & 0.13& \textcolor{blue}{578.27} &  \textcolor{blue}{$\mathbf{0.00}$}& -2.75 & 0.22& \textcolor{red}{-5.82} &  \textcolor{red}{$\mathbf{0.00}$}\\
 & 2010 & 0.00& -1.52 & 0.22& \textcolor{blue}{842.87} &  \textcolor{blue}{$\mathbf{0.00}$}& -0.64 & 0.80& \textcolor{red}{-5.19} &  \textcolor{red}{$\mathbf{0.00}$}\\
\hline
Univeristy of Vermont & 1974 & 0.00& -0.39 & 0.54& \textcolor{blue}{20.93} &  \textcolor{blue}{$\mathbf{0.00}$}& -0.49 & 0.54& -1.17 & 0.13\\
 & 1980 & 0.00& -0.72 & 0.41& \textcolor{blue}{82.27} &  \textcolor{blue}{$\mathbf{0.00}$}& -0.81 & 0.47& \textcolor{red}{-1.82} &  \textcolor{red}{$\mathbf{0.06}$}\\
 & 1990 & 0.00& -0.94 & 0.36& \textcolor{blue}{23.05} &  \textcolor{blue}{$\mathbf{0.01}$}& -1.11 & 0.34& \textcolor{red}{-1.79} &  \textcolor{red}{$\mathbf{0.06}$}\\
 & 2000 & 0.00& -0.65 & 0.45& \textcolor{blue}{30.59} &  \textcolor{blue}{$\mathbf{0.00}$}& -0.78 & 0.44& \textcolor{red}{-1.52} &  \textcolor{red}{$\mathbf{0.08}$}\\
 & 2010 & 0.00& -$\infty$ & \textcolor{red}{$\mathbf{0.00}$}& \textcolor{blue}{7.75} &  \textcolor{blue}{$\mathbf{0.02}$}& 0.39 & 0.34& -0.00 & 0.94\\
\hline
United Way, Chittendon County & 2004 & 0.00& -0.46 & 0.47& \textcolor{blue}{28.75} &  \textcolor{blue}{$\mathbf{0.00}$}& -0.53 & 0.55& -1.29 & 0.11\\
 & 2005 & 0.00& -0.08 & 0.77& \textcolor{blue}{54.69} &  \textcolor{blue}{$\mathbf{0.00}$}& 0.36 & 0.74& -0.69 & 0.24\\
 & 2006 & 0.00& -0.12 & 0.71& \textcolor{blue}{68.71} &  \textcolor{blue}{$\mathbf{0.00}$}& 0.44 & 0.71& -0.85 & 0.19\\
 & 2007 & 0.00& -0.61 & 0.43& \textcolor{blue}{48.21} &  \textcolor{blue}{$\mathbf{0.00}$}& -0.65 & 0.57& \textcolor{red}{-1.64} &  \textcolor{red}{$\mathbf{0.07}$}\\
 & 2008 & 0.00& -0.13 & 0.72& \textcolor{blue}{46.52} &  \textcolor{blue}{$\mathbf{0.00}$}& 0.14 & 0.90& -0.71 & 0.23\\
 & 2009 & 0.00& -0.35 & 0.55& \textcolor{blue}{48.39} &  \textcolor{blue}{$\mathbf{0.00}$}& -0.28 & 0.80& -1.15 & 0.13\\
 & 2010 & 0.00& -0.32 & 0.58& \textcolor{blue}{35.25} &  \textcolor{blue}{$\mathbf{0.00}$}& -0.30 & 0.77& -0.90 & 0.18\\
\hline
ECHO Science Museum & 2005 & 0.00& -2.47 & 0.25& \textcolor{blue}{31.43} &  \textcolor{blue}{$\mathbf{0.04}$}& -3.04 & 0.21& \textcolor{red}{-3.56} &  \textcolor{red}{$\mathbf{0.01}$}\\
 & 2006 & 0.00& -0.20 & 0.69& 1.42 & 0.57& -0.28 & 0.68& -0.53 & 0.30\\
 & 2007 & 0.00& -$\infty$ & \textcolor{red}{$\mathbf{0.00}$}& \textcolor{blue}{4.56} &  \textcolor{blue}{$\mathbf{0.03}$}& 0.20 & 0.35& 0.00 & 1.00\\
 & 2008 & 0.00& -$\infty$ & \textcolor{red}{$\mathbf{0.00}$}& \textcolor{blue}{4.28} &  \textcolor{blue}{$\mathbf{0.00}$}& 0.29 & 0.19& 0.00 & 1.00\\
 & 2009 & 0.00& -0.87 & 0.47& \textcolor{blue}{31.48} &  \textcolor{blue}{$\mathbf{0.01}$}& -1.23 & 0.44& \textcolor{red}{-2.51} &  \textcolor{red}{$\mathbf{0.03}$}\\
\hline
Flynn Theater & 2006 & 0.00& -0.52 & 0.46& \textcolor{blue}{272.93} &  \textcolor{blue}{$\mathbf{0.00}$}& 0.32 & 0.87& \textcolor{red}{-2.80} &  \textcolor{red}{$\mathbf{0.02}$}\\
 & 2007 & 0.00& -0.06 & 0.80& 4.53 & 0.14& -0.08 & 0.86& -0.26 & 0.47\\
 & 2008 & 0.00& -0.56 & 0.45& \textcolor{blue}{303.73} &  \textcolor{blue}{$\mathbf{0.00}$}& 0.38 & 0.86& \textcolor{red}{-3.35} &  \textcolor{red}{$\mathbf{0.01}$}\\
 & 2009 & 0.00& -0.25 & 0.63& \textcolor{blue}{281.34} &  \textcolor{blue}{$\mathbf{0.00}$}& 1.11 & 0.59& \textcolor{red}{-2.16} &  \textcolor{red}{$\mathbf{0.04}$}\\
 & 2010 & 0.00& \textcolor{red}{-3.96} &  \textcolor{red}{$\mathbf{0.07}$}& \textcolor{blue}{129.19} &  \textcolor{blue}{$\mathbf{0.00}$}& \textcolor{red}{-4.61} &  \textcolor{red}{$\mathbf{0.06}$}& \textcolor{red}{-6.78} &  \textcolor{red}{$\mathbf{0.00}$}\\
\hline
\end{tabular}
\caption{The results of the Likelihood-Ratio and its associated {\bf p}-value are reported for different distributions. Here, positive values lend support to the Power Law and negative values to the other stated distribution. The significance of the {\bf LR} is {\bf p}, where low values of {\bf p} indicate a trustworthy {\bf LR}. Values for which $\mathbf{ p} < 0.05$ are bolded.}
\label{table:fittest}
\end{table*}

\begin{table*}
\begin{tabular}{|lc|rrr|cl|cc|cl|}
\hline
{\bf Name}~~~~~~~~~~~~~~~~~ & {\bf Year} & $~~~~~~~~~~~\mathbf{ \langle x \rangle}$ & $~~~~~~~~~~~~~\mathbf{ \sigma}$ & $~~~~~~\mathbf{x_\text{max}}$ &  $~~~~~~\mathbf{\gamma_\text{MLE}}~~~~~~$ & $\mathbf{x_\text{min} }~~~~~$ & ~~~{\bf D} ~~~ & ~~~{\bf p}~~~ & ~~~~~~~$\mathbf{\gamma_\text{LS}}$~~~~~~~ & {\bf Range}\\ 
\hline
Romney & 2010 & 94456.52 & 336755.59 & 1670000 & 2.09 $\pm$ 0.25 & 10000 & 0.11 & 0.00 & 1.65 $\pm$ 0.13 & 1 to 23\\
\hline
McCain & 2006 & 11037.59 & 15077.01 & 50500 & 1.88 $\pm$ 0.28 & 4000 & 0.20 & 0.00 & 1.47 $\pm$ 0.12 & 1 to 17\\
\hline
Obama & 2011 & 4413.59 & 18330.81 & 117130 & 3.16 $\pm$ 0.39 & 1000 & 0.13 & 0.00 & 1.96 $\pm$ 0.26 & 1 to 39\\
\hline
HW Bush & 1990 & 749.94 & 1116.48 & 5521 & 3.35 $\pm$ 0.61 & 1000 & 0.17 & 0.00 & 1.79 $\pm$ 0.19 & 1 to 52\\
\hline
Clinton & 1992 & 2083.75 & 3117.79 & 10220 & 2.06 $\pm$ 0.43 & 550 & 0.15 & 0.50 & 1.69 $\pm$ 0.12 & 1 to 8\\
\hline
Nixon & 1972 & 73.75 & 73.77 & 200 & 2.16 $\pm$ 0.58 & 20 & 0.18 & 1.00 & 1.58 $\pm$ 0.11 & 1 to 4\\
\hline
\end{tabular}
\caption{Summary statistics of all of the Presidential donation data.}
\label{table:presfitstats}
\end{table*}
\begin{table*}
\begin{tabular}{|lc|c|cc|cc|cc|cc|}
\hline
 &  & & \multicolumn{2}{|c|}{{\bf Log-Normal }} & \multicolumn{2}{|c|}{{\bf Exponential }} & \multicolumn{2}{|c|}{{\bf Stretched Exp. }} & \multicolumn{2}{|c|}{{\bf Cutoff Power Law}} \\ 
{\bf Name}~~~~~~~~~~~~~~~~ & {\bf Year} & {\bf ~~p~~} & {\bf ~~LR~~} & {\bf ~~p~~}  &  {\bf ~~LR~~} & {\bf ~~p~~} & {\bf ~~LR~~} & {\bf ~~p~~} & {\bf ~~LR~~} & {\bf ~~p~~}\\ 
\hline
Romney & 2010 & 0.00&     -$\infty$ & \textcolor{red}{$\mathbf{0.00}$}& \textcolor{blue}{   28.63} &  \textcolor{blue}{$\mathbf{0.00}$}&     0.70 & 0.15&     0.00 & 1.00\\
\hline
McCain & 2006 & 0.00&    -0.39 & 0.60&     0.10 & 0.96&    -0.56 & 0.58&    -0.75 & 0.22\\
\hline
Obama & 2011 & 0.00&     -$\infty$ & \textcolor{red}{$\mathbf{0.00}$}& \textcolor{blue}{   55.53} &  \textcolor{blue}{$\mathbf{0.00}$}& \textcolor{blue}{    2.67} &  \textcolor{blue}{$\mathbf{0.01}$}&     0.00 & 1.00\\
\hline
HW Bush & 1990 & 0.00&     -$\infty$ & \textcolor{red}{$\mathbf{0.00}$}& \textcolor{blue}{    5.27} &  \textcolor{blue}{$\mathbf{0.01}$}& \textcolor{blue}{    0.55} &  \textcolor{blue}{$\mathbf{0.09}$}&     0.00 & 1.00\\
\hline
Clinton & 1992 & 0.50&     -$\infty$ & \textcolor{red}{$\mathbf{0.00}$}&     2.58 & 0.15&     0.06 & 0.78&    -0.02 & 0.83\\
\hline
Nixon & 1972 & 1.00&     -$\infty$ & \textcolor{red}{$\mathbf{0.00}$}&     1.11 & 0.41&     0.03 & 0.85&    -0.03 & 0.82\\
\hline
\end{tabular}
\caption{The results of the Likelihood-Ratio and its associated {\bf p}-value are reported for the presidential candidates.}
\label{table:presfittest}
\end{table*}


\begin{thebibliography}{16}
\expandafter\ifx\csname natexlab\endcsname\relax\def\natexlab#1{#1}\fi
\expandafter\ifx\csname bibnamefont\endcsname\relax
  \def\bibnamefont#1{#1}\fi
\expandafter\ifx\csname bibfnamefont\endcsname\relax
  \def\bibfnamefont#1{#1}\fi
\expandafter\ifx\csname citenamefont\endcsname\relax
  \def\citenamefont#1{#1}\fi
\expandafter\ifx\csname url\endcsname\relax
  \def\url#1{\texttt{#1}}\fi
\expandafter\ifx\csname urlprefix\endcsname\relax\def\urlprefix{URL }\fi
\providecommand{\bibinfo}[2]{#2}
\providecommand{\eprint}[2][]{\url{#2}}

\bibitem[{giv()}]{givingusa2011a}
\bibinfo{note}{Giving USA Foundation (2011). Giving USA 2011: The Annual Report
  on Philanthropy for the Year 2010. Retrieved from www.givingusareports.org.}

\bibitem[{\citenamefont{Clementi and Gallegati}(2005)}]{clementi2005power}
\bibinfo{author}{\bibfnamefont{F.}~\bibnamefont{Clementi}} \bibnamefont{and}
  \bibinfo{author}{\bibfnamefont{M.}~\bibnamefont{Gallegati}},
  \bibinfo{journal}{Physica A: Statistical Mechanics and its Applications}
  \textbf{\bibinfo{volume}{350}}, \bibinfo{pages}{427} (\bibinfo{year}{2005}).

\bibitem[{\citenamefont{Nirei and Souma}(2007)}]{nirei2007two}
\bibinfo{author}{\bibfnamefont{M.}~\bibnamefont{Nirei}} \bibnamefont{and}
  \bibinfo{author}{\bibfnamefont{W.}~\bibnamefont{Souma}},
  \bibinfo{journal}{Review of Income and Wealth} \textbf{\bibinfo{volume}{53}},
  \bibinfo{pages}{440} (\bibinfo{year}{2007}).

\bibitem[{\citenamefont{Wu et~al.}(2011)\citenamefont{Wu, Guo, Chen, and
  Wang}}]{wu2011a}
\bibinfo{author}{\bibfnamefont{Y.}~\bibnamefont{Wu}},
  \bibinfo{author}{\bibfnamefont{J.}~\bibnamefont{Guo}},
  \bibinfo{author}{\bibfnamefont{Q.}~\bibnamefont{Chen}}, \bibnamefont{and}
  \bibinfo{author}{\bibfnamefont{Y.}~\bibnamefont{Wang}},
  \bibinfo{journal}{Physica A} \textbf{\bibinfo{volume}{390}},
  \bibinfo{pages}{4325} (\bibinfo{year}{2011}).

\bibitem[{\citenamefont{Chen et~al.}(2009)\citenamefont{Chen, Wang., and
  Wang.}}]{chen2009a}
\bibinfo{author}{\bibfnamefont{Q.}~\bibnamefont{Chen}},
  \bibinfo{author}{\bibfnamefont{C.}~\bibnamefont{Wang.}}, \bibnamefont{and}
  \bibinfo{author}{\bibfnamefont{Y.}~\bibnamefont{Wang.}},
  \bibinfo{journal}{Europhysics Letters} \textbf{\bibinfo{volume}{88}},
  \bibinfo{pages}{38001} (\bibinfo{year}{2009}).

\bibitem[{\citenamefont{Pierpoint and Wilkerson}(1998)}]{pierpoint1998a}
\bibinfo{author}{\bibfnamefont{R.}~\bibnamefont{Pierpoint}} \bibnamefont{and}
  \bibinfo{author}{\bibfnamefont{G.~S.} \bibnamefont{Wilkerson}},
  \bibinfo{journal}{New directions for Philanthropic Fundraising}
  \textbf{\bibinfo{volume}{21}}, \bibinfo{pages}{61} (\bibinfo{year}{1998}).

\bibitem[{\citenamefont{Zipf}(1949)}]{zipf1949a}
\bibinfo{author}{\bibfnamefont{G.~K.} \bibnamefont{Zipf}},
  \emph{\bibinfo{title}{Human Behaviour and the Principle of Least-Effort}}
  (\bibinfo{publisher}{Addison-Wesley}, \bibinfo{address}{Cambridge, MA},
  \bibinfo{year}{1949}).

\bibitem[{\citenamefont{Clauset et~al.}(2009)\citenamefont{Clauset, Shalizi,
  and Newman}}]{clauset2009b}
\bibinfo{author}{\bibfnamefont{A.}~\bibnamefont{Clauset}},
  \bibinfo{author}{\bibfnamefont{C.~R.} \bibnamefont{Shalizi}},
  \bibnamefont{and} \bibinfo{author}{\bibfnamefont{M.~E.~J.}
  \bibnamefont{Newman}}, \bibinfo{journal}{SIAM Review}
  \textbf{\bibinfo{volume}{51}}, \bibinfo{pages}{661} (\bibinfo{year}{2009}).

\bibitem[{\citenamefont{Dodds et~al.}(2001)\citenamefont{Dodds, Rothman, and
  Weitz}}]{dodds2001d}
\bibinfo{author}{\bibfnamefont{P.~S.} \bibnamefont{Dodds}},
  \bibinfo{author}{\bibfnamefont{D.~H.} \bibnamefont{Rothman}},
  \bibnamefont{and} \bibinfo{author}{\bibfnamefont{J.~S.} \bibnamefont{Weitz}},
  \bibinfo{journal}{Journal of Theoretical Biology}
  \textbf{\bibinfo{volume}{209}}, \bibinfo{pages}{9} (\bibinfo{year}{2001}).

\bibitem[{\citenamefont{on~Philanthropy~at
  Indiana~University}(2007)}]{copii2007a}
\bibinfo{author}{\bibfnamefont{C.}~\bibnamefont{on~Philanthropy~at
  Indiana~University}}, \emph{\bibinfo{title}{Patterns of household charitable
  giving by income group}}, \bibinfo{howpublished}{Prepared for Google}
  (\bibinfo{year}{2007}).

\bibitem[{\citenamefont{Goldenfeld}(1992)}]{goldenfeld1992a}
\bibinfo{author}{\bibfnamefont{N.}~\bibnamefont{Goldenfeld}},
  \emph{\bibinfo{title}{Lectures on Phase Transitions and the Renormalization
  Group}}, vol.~\bibinfo{volume}{85} of \emph{\bibinfo{series}{Frontiers in
  Physics}} (\bibinfo{publisher}{Addison-Wesley}, \bibinfo{address}{Reading,
  Massachusetts}, \bibinfo{year}{1992}).

\bibitem[{\citenamefont{Simon}(1955)}]{simon1955a}
\bibinfo{author}{\bibfnamefont{H.~A.} \bibnamefont{Simon}},
  \bibinfo{journal}{Biometrika} \textbf{\bibinfo{volume}{42}},
  \bibinfo{pages}{425} (\bibinfo{year}{1955}).

\bibitem[{\citenamefont{Dove}(2000)}]{dove2000a}
\bibinfo{author}{\bibfnamefont{K.~E.} \bibnamefont{Dove}},
  \emph{\bibinfo{title}{Conducting a Successful Capital Campaign}}
  (\bibinfo{publisher}{Jassey-Bass}, \bibinfo{address}{San Francisco},
  \bibinfo{year}{2000}), p.~\bibinfo{pages}{72}, \bibinfo{edition}{2nd} ed.

\bibitem[{gra(2011)}]{graham2011b}
 (\bibinfo{year}{2011}), \bibinfo{note}{{C}hristine {G}raham, personal
  communication}.

\bibitem[{\citenamefont{Newman}(2005)}]{newman2005b}
\bibinfo{author}{\bibfnamefont{M.~E.~J.} \bibnamefont{Newman}},
  \bibinfo{journal}{Contemporary Physics} \textbf{\bibinfo{volume}{46}},
  \bibinfo{pages}{323} (\bibinfo{year}{2005}).

\bibitem[{\citenamefont{Alstott}(2012)}]{alstott}
\bibinfo{author}{\bibfnamefont{J.}~\bibnamefont{Alstott}},
  \bibinfo{journal}{Web address: pypi. python. org/pypi/powerlaw}
  (\bibinfo{year}{2012}).

\end{thebibliography}
\end{document}